\documentclass{aa}  

\usepackage{graphicx}
\usepackage{txfonts}
\usepackage{natbib}
\usepackage[citecolor=blue,colorlinks=true,linkcolor=blue,urlcolor=blue]{hyperref}
\usepackage[export]{adjustbox}
\usepackage{tablefootnote}

\begin{document} 

\title{Time-evolving Diagnostic of the Ionized Absorbers in NGC 4051. II. High-throughput Time-resolved Spectroscopy}

\author{Alfredo Luminari
\inst{1,2}\thanks{\email{alfredo.luminari@inaf.it}}
\and Fabrizio Nicastro \inst{1}
\and Roberto Serafinelli \inst{3,1}
\and Riccardo Middei \inst{1}
\and Yair Krongold \inst{4}
\and Elias Kammoun \inst{5}
\and Luigi Piro \inst{2}
\and Francesca Diaferia \inst{6,1}
}

\institute{
INAF – Osservatorio Astronomico di Roma, Via Frascati 33, 00078 Monteporzio, Italy
\and
INAF – Istituto di Astrofisica e Planetologia Spaziali, Via del Fosso del Caveliere 100, 00133 Roma, Italy
\and
Instituto de Estudios Astrofísicos, Universidad Diego Portales, Avenida Ejército Libertador 441, Santiago, Chile
\and
Instituto de Astronomía, Universidad Nacional Autónoma de México, Ciudad Universitaria, Ciudad de México 04510, México
\and
Cahill Center for Astronomy \& Astrophysics, California Institute of Technology, 1216 East California Boulevard, Pasadena, CA
\and Dipartimento di Fisica, Sapienza Università di Roma, P.le Aldo Moro, 00100 Roma, Italy
}
\date{Received 29 April 2026 / Accepted 21 July 2026}

\abstract
{Active Galactic Nuclei (AGNs) are one of the most powerful sources in the Universe. The energy liberated by the accretion onto the central supermassive black hole can strongly impact the surrounding environment, up to the host galaxy and beyond. X-ray spectroscopy is a powerful probe of the AGN structure and of the nuclear scale outflows. Notwithstanding their ubiquitous presence, such energetic outflows are so far poorly characterised. This is mainly due to the degeneracy between their number density $n_e$ and radial location $r$, intrinsic in the photoionisation equilibrium models which are usually employed to fit the observations. This degeneracy prevents a self-consistent determination of the gas mass and energy rates and, therefore, it does not allow to assess the potential role of these outflows in transporting the accretion-liberated energy outwards.}
{We analyse a joint XMM-Newton EPIC-pn and NuSTAR observation of NGC 4051, one of the brightest and most variable local AGN. X-ray observations in the last 25 years revealed three main warm ionised absorbers in NGC 4051, together with emission lines coming from different accretion disc radii. The high flux allows to perform time-resolved spectroscopy and, thus, to study the evolution of the absorbers following the variability of the incident ionising continuum. Since the timescale of the gas ionisation variability depends on its number density, constraining it allows to break the density-distance degeneracy, finally enabling a precise determination of the key gas properties.}
{We employ the Time-Evolving PhotoIonisation Device (TEPID) to model the temporal evolution of the outflows. We split the observations in 37 time-resolved spectra, each few kiloseconds long (total duration 160 ksec), and we fit them jointly with the time-resolved ionised spectra predicted by TEPID. 
The column densities and the line-of-sight velocities were derived as well through the analysis of the simultaneous RGS grating spectra in our previous paper (Serafinelli et al. 2025).}
{Thanks to our time-resolved analysis, we fully constrain the number density, $n_e=10^7 cm^{-3}$, and the distance, $r=10^4$ gravitational radii ($r_G$), of the absorber with the highest opacity and with intermediate ionisation. This distance is the same of the Broad Line Region observed in the optical and UV bands and of the soft X-ray emission lines. We obtain upper and lower limits for the other two absorbers. The fastest and most ionised one is at $r<600 r_G$, where the broad component of the Fe K$\alpha$ line at 6.4 keV also originates. The slowest and least ionised absorber is at $r \geq 3.8 \cdot 10^5 r_G$, the same distance of the cold torus and the narrow Fe K$\alpha$. Finally, the total energy outflow rate is below $10^{-4}$ the bolometric luminosity of the AGN $L_{bol}$, ruling out a meaningful mechanical impact on the host galaxy.
These results demonstrate the potential held by time-evolving photoionisation in delivering a clear view of the nuclear outflows and their connection with the AGN structure.}
{}

\authorrunning{A. Luminari et al.}
\titlerunning{Time-evolving Diagnostic of the Ionized Absorbers in NGC 4051. II.}

\maketitle

\section{Introduction \label{sec:intro}}
Active Galactic Nuclei (AGNs) are the most powerful steady luminosity sources in the Universe. The energy liberated by the accretion of matter onto the supermassive black hole at the centre of a galaxy is emitted at all wavelengths, with the bulk between the optical and the X-ray. Spectroscopy is a powerful tool to investigate the inner structure and the accretion and ejection mechanisms, with the X-ray (and the UV) band probing the innermost scale. All the main components can be detected in the X-ray, i.e. the central, compact "corona" emitting the primary radiation \citep{hm91,2018A&A...614A..37T,2024A&A...690A.145S}, the gravitationally-bound structures (such as the virialised Broad Lines and the reflection from the accretion disc; \citealp{2024ApJ...973L..25X,2026arXiv260216252B,2025ApJ...994L..13K} for recent XRISM/Resolve results) and the outflowing matter (\citealp{2025ApJ...993L..53N,2025Natur.641.1132X,2025A&A...699A.228M}; see  \citealp{2024ApJS..274....8Y} for a comprehensive review). Outflows are primarily detected via absorption from photo- (or, eventually, collisional-) ionised gas and are divided in two broad categories. The "Warm Absorbers" (WAs) are observed in roughly 50\% of AGNs \citep{2012ApJ...745..107W}, are characterised by moderately ionised metals, such as O VII, N VI, Ne IX, Mg XII, Fe XX, with the bulk of the opacity in the soft X-ray band (E$\leq$ 2 KeV), moderate line-of-sight velocity ($v_{out} \lesssim  few$ thousands km\ s$^{-1}$) and Hydrogen-equivalent column density $N_H < 10^{23} cm^{-2}$ (see e.g. \citealp{2002ApJ...574..643K}). On the other hand, the "Ultra-Fast Outflows" have mildly relativistic $v_{out}$, up to 0.1-0.3 times the speed of light $c$, thicker columns, $N_H \approx 10^{24} cm^{-2}$, and higher ionisation, so that the only bound electrons are often in He- and H-like Iron \citep{fiore23,matzeu23}. The two populations are not neatly divided and an increasing number of intermediate cases is being observed (see e.g. \citealp{krongold21,2015ApJ...813L..39L,2026arXiv260424899M}).

Despite their ubiquitous presence and their potential to efficiently transfer the accretion-liberated energy at galaxy scales \citep{Fiore17}, the physical origins of these outflows and their connection with the accretion flow is still uncertain. This is mainly due to the difficulty in constraining their number density and radial location and, thus, to reliably estimate the outflowing mass and its ability to escape the gravitational potential well.
In the time-equilibrium ionisation approximation, which is usually adopted to model and fit the observations, the gas is assumed to have had enough time to get in equilibrium with its incident ionising flux. When so, its physical status is uniquely parametrised by $U=Q_{ion}/4 \pi c n_H r^2$, where $Q_{ion}$ is the ionising photon luminosity and $r, n_H$ are the distance of the absorber from the luminosity source and its hydrogen number density, respectively. This implies a direct degeneracy between $n_H$ and $r$.

A promising and so far poorly explored strategy to break this degeneracy is to consider the so-called "equilibration timescale", $t_{eq}$, over which the gas readjusts following intrinsic luminosity variations. It can be shown that, to a first order, this timescale is inversely proportional to the gas electron number density, $t_{eq} \propto 1/n_e$ (\citealp{kk95,nicastro99}; $n_e \sim 1.2 n_H$ for solar-like metallicity). Constraining $n_e$, together with the ionisation parameter $U$, allows then to derive the gas radial distance as:
\begin{equation}
r=\sqrt{Q_{ion}/4 \pi U n_H c}
\label{eq_r}
\end{equation}
Time-evolving ionisation is thus a promising channel to constrain the gas location, density and, thus, to help derive its geometry, kinematics and energetic. 
Several time-evolving ionisation codes have become available recently, such as TPHO \citep{rogantini22,kosec24}, the XSTAR-derived one presented in \cite{sadaula23,2025ApJ...992..182S} and the Time-Evolving PhotoIonisation Device (TEPID; \citealp{luminari23b,2026NatAs.tmp...42T}) \footnote{See also \cite{garcia13,2002ApJ...580..261P} for previous time-evolving models of low density gaseous nebulae and Gamma-Ray Bursts afterglows, respectively.}. TEPID follows the (out-of-equilibrium) evolution of the ionisation and the physics of a gas illuminated by a time-variable ionising source. The predicted ionisation as a function of time (or, better, the time-resolved ionic columns) can be directly fitted to observed spectra via a customised version of the PHASE \citep{krongold03} spectral synthesis package. 

In this paper we focus on NGC 4051, a nearby galaxy (z=0.00234, \citealp{devaucouleurs91}) with one of the brightest and together most variable active galactic nucleus at its centre. Spectroscopy, mainly in the X-ray and UV-optical bands, have demonstrated the widespread presence of ionised nuclear absorbers which are persistent over the last $\sim$ 25 years \citep{nicastro99,2001ApJ...557....2C, 2011MNRAS.414.1965L,2012ApJ...751...84K,2019ApJ...879..102P}.  Thanks to the high flux of NGC 4051, X-ray variability can be exploited to study the ionisation variation of its WAs. The first systematic study on this topic was \cite{nicastro99}, followed by \cite{kne07} (hereafter K07), which exploited $t_{eq}$ to constrain the number density, and thus the location, of the two main absorbers. 

This paper follows \cite{serafinelli25} (hereafter Paper I) in the analysis of a joint 2018 observation. NGC 4051 was observed with two 80 ksec-long pointings with XMM-Newton \citep{jansen01}, embedded in a longer, simultaneous NuSTAR \citep{nustar} observation. Paper I analysed the two 80 ksec-long time-averaged RGS spectra. They found evidence of three distinct WAs, dubbed Low- and High-Ionisation Phase (LIP, HIP) and High-Velocity and -Ionisation Phase (HVIP). They have increasing ionisation parameter and column density, with $-1.0 \leq log(U) \leq 2.3$ and $20.2 \leq log(N_H/cm^{-2}) \leq 21.43$. The significant X-ray variability during the observations allowed to apply TEPID and, thus, to constrain density and location of the HIP to around $5 \cdot 10^7 cm^{-3}$ and $3.7 \cdot 10^{-4} pc$, marginally consistent with the location of the Broad Line Region.
In this Paper, instead, we study infra-observation spectral variability by analysing the XMM-Newton EPIC-pn \citep{struder01} and NuSTAR \citep{nustar} observations in a time-resolved fashion. 
In Sect. \ref{sec:Data_reduction} we detail the data reduction. TEPID modelling is discussed in Sect. \ref{TEPID}, while its application to the data is presented in Sect. \ref{Data Analysis}. Discussions and Conclusions are in Sect. \ref{sec:Discussion} and \ref{sec:Conclusion}.

\section{Data reduction \label{sec:Data_reduction}}
\subsection{XMM-Newton PN}
Our dataset is composed by two \textit{XMM-Newton} observations (obsID 0830430201 and 0830430801, Obs1 and Obs2 hereafter), each one $\sim 80$ ksec long and separated by 90 ksec. We processed the EPIC-pn dataset using the XMM SAS (version 21.1.0)\footnote{\url{https://xmm-tools.cosmos.esa.int/external/xmm_user_support/documentation/sas_usg/USG/}} via the standard calibration commands, \texttt{cifbuild} and \texttt{odfingest}, and creating the calibrated files with \texttt{epproc}. 
We cleaned the event files using the SAS filters \texttt{XMMEA\_EP} and \texttt{PATTERN==0}. This ensures that only the events with the best spectral quality are included in the analysis. Figure \ref{fig:LC_XMM_Nustar}, top panel, shows the XMM-Newton 2-10 keV and NuSTAR 3-79 keV (see below) lightcurves.

To optimally resolve the spectral variations, we subsequently divide Obs1 in temporal bins of $5 \cdot 10^4$ counts in the 0.5 to 10 keV band (for a total of 14), while Obs2 in bins of $2 \cdot 10^4$ counts, since it had slightly lower flux (for a total of 23). Figure \ref{fig:LC_XMM_Nustar}, bottom panel, shows the time and the flux of each bin. Their duration is between 1700 and 5700 seconds.

\begin{figure}[]
\centering
\includegraphics[width=\columnwidth]{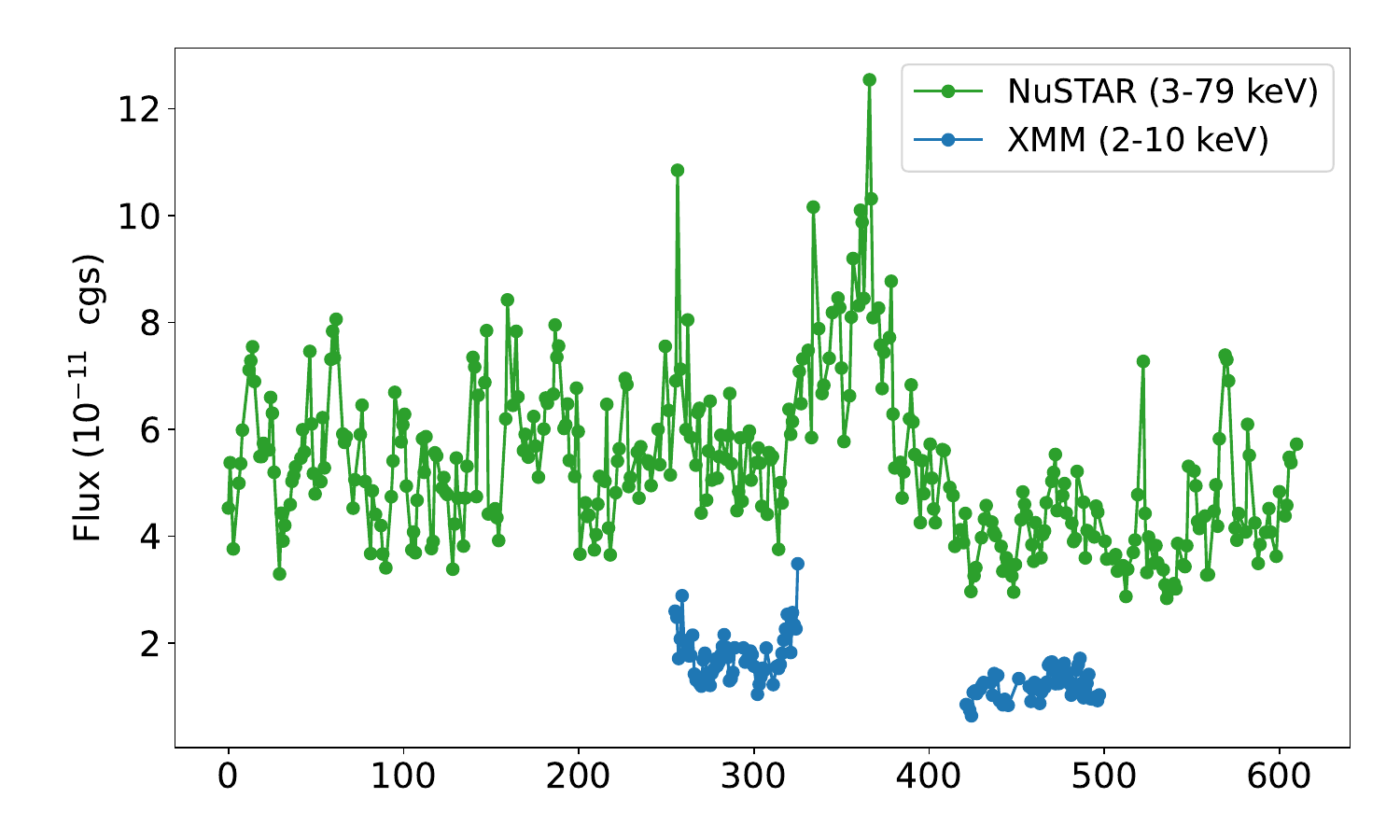} \\
\includegraphics[width=\columnwidth]{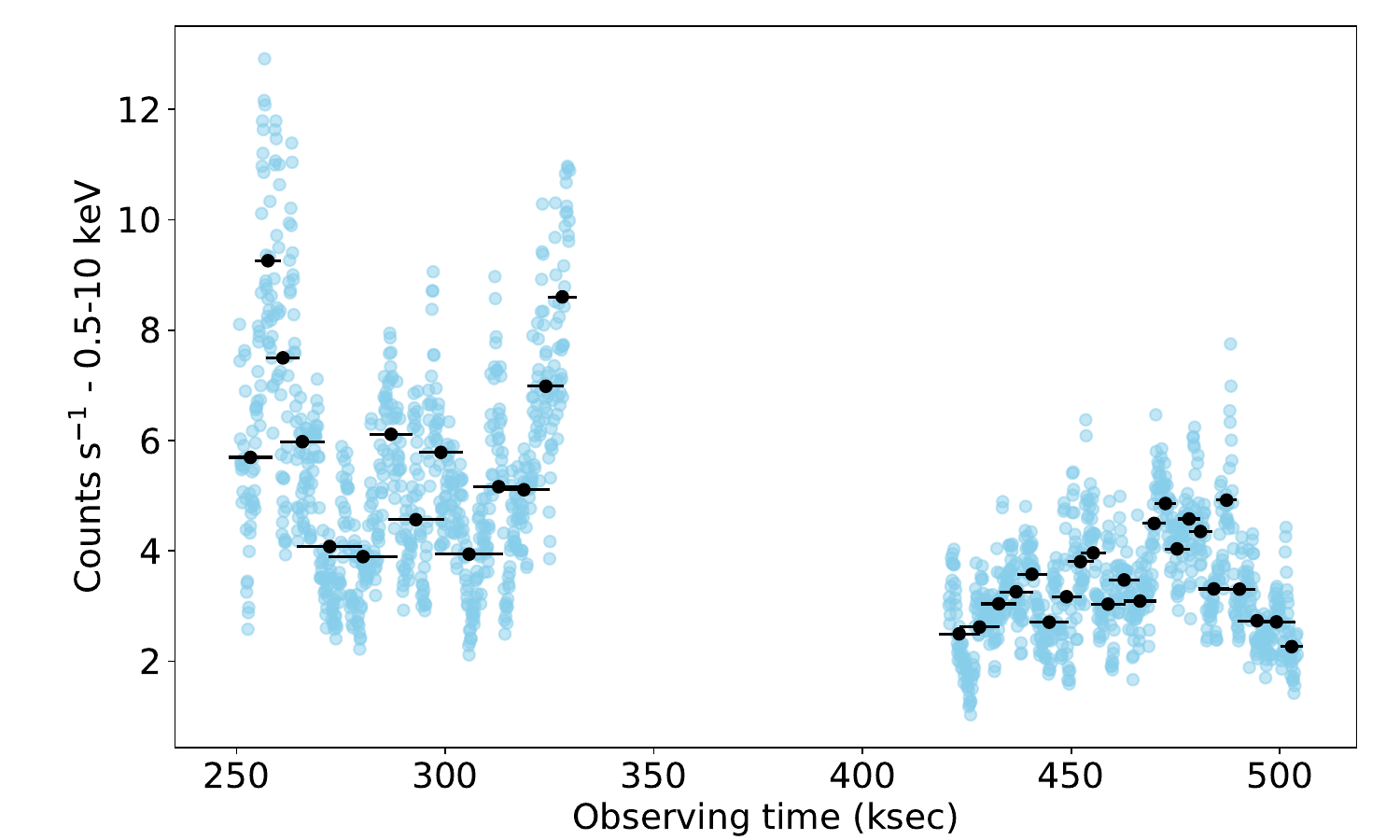} 
\caption{Top: NGC 4051 lightcurve in the 2.10 keV range with XMM-Newton (blue) and in the 3-79 keV range with NuSTAR (green). The time is from the start of the NuSTAR observation. Bottom: XMM-Newton 0.5-10 keV lightcurve (blue) and the average flux of the time bins (black).} 
\label{fig:LC_XMM_Nustar}
\end{figure}

\subsection{NuSTAR}
\textit{Nustar} observed NGC 4051 (obsID 60401009002) for 600 ksec, starting 247 ksec before the XMM observations and fully encompassing it. We create the event calibrated files with the \texttt{nupipeline} reduction tool, provided in the NuSTARDAS\footnote{\url{https://heasarc.gsfc.nasa.gov/docs/nustar/analysis/}} reduction software suite (version 1.10). We extract FPMA and FPMB spectra both simultaneous with XMM Obs1 and Obs2. Then, we extract 37 time-resolved spectra corresponding to the XMM-Newton time bins. Such spectra will prove useful to probe the high-energy end of the continuum and constrain the reflection component.

\section{The Time-Evolving PhotoIonisation Device (TEPID) model}
\label{TEPID}

\subsection{Code updates}
TEPID has been significantly updated since the first version presented in \cite{luminari23b}, mainly concerning the radiative transfer. While the previous version only accounted for photoelectric absorption along the gas column, the code now also computes (and includes in the radiative transfer) the radiative recombination continuum (RRC) and the free-free (brehmsstrahlung) emission spectra.

RRC spectrum is computed (in units of $cm^{-2} s^{-1} keV^{-1}$) as (see e.g. \citealp{of06}):

\begin{equation}
\begin{split}
& RRC = \\ & \sum_{X,i} \alpha_{rec}(X,i)n_e n_{ion} (X,i) \Big( \frac{1-P_c(\tau)}{\kappa_0+\kappa_n (X,i)} \Big) \frac{2 \sqrt{h\nu-h \nu_{X,i}}}{\sqrt{\pi} (kT)^{3/2}} \cdot e^{-\frac{h\nu-h \nu_{X,i}}{kT}} \\
\end{split}
\end{equation}

where the sum extends over all the elements $X$ and ionic levels $i$ and for all the frequencies above the photoionisation threshold energy, i.e. $h\nu-h \nu_{X,i}>0$. $P_c(\tau)=\frac{1-e^{-\tau_{\nu}}}{\tau_{\nu}}$ is the frequency-dependent escape probability for the given ion $X,i$. $\kappa_0=n_e \cdot \sigma_T$ is the Thompson opacity, while $\kappa_n(X,i)=n_{ion} \cdot \sigma_{\nu}^{X,i}$ is the photoelectric opacity ($\sigma_{\nu}^{X,i}$ is the frequency-dependent photoionisation rate). Both are in units of $cm^{-1}$. $\alpha_{rec} (X,i)$ is the radiative recombination rate of the ion.

Free-free emission spectrum is computed as in Eq. 5.14b from \cite{rl79}. The emissivity (in $keV\ s^{-1}\ cm^{-3}\ keV^{-1}$) is given by:
\begin{equation}
\epsilon_{\nu}^{ff}= \sum_{X,i}\ 1.02 \cdot 10^{-11}Z^2 n_e n_i T^{-1/2} e^{-h \nu /kT} \overline{g}_{ff}
\end{equation}
where $Z$ is the number of unbound electrons for the ion $X,i$ and $\overline{g}_{ff}$ are the temperature-averaged Gaunt factors. The spectrum (in units of $cm^{-2} s^{-1} keV^{-1}$) is the integral of $\epsilon_{\nu}^{ff}$ over the wind column:
\begin{equation}
E_{\nu} = \int \frac{\epsilon_{\nu}^{ff}}{h \nu} dr 
\end{equation}

TEPID now also includes three-body recombination. The rates are derived from the respective collisional ionisation rates as in Eq. 5 in \cite{Bautista2001} \footnote{We set the weight of the starting and arrival levels =1 since at the moment TEPID assumes all the electrons to lye in the ground state}. Finally, the description of the variability of the incident luminosity has been significantly improved. In the former version, the Spectral Energy Distribution (SED) varied rigidly in time, with the same variation throughout the entire TEPID energy range ($10^{-5} - 10^2$ keV). Now instead, the SED can vary arbitrarily from one time bin to the following. As we will show in Sect. \ref{4051 sims} below, this allows to account for different variability at different energies, as is the case in AGNs, where variability is typically higher in the X-rays than at lower energies.

\begin{figure}
\centering
\includegraphics[width=\columnwidth]{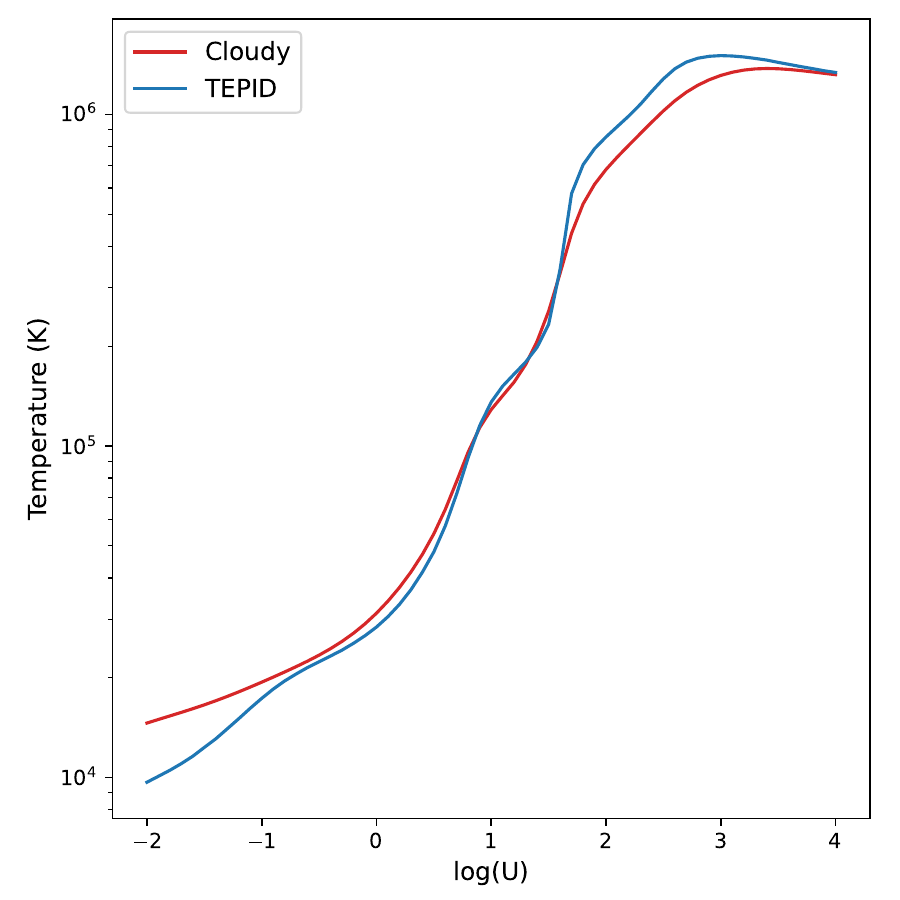}
\caption{Equilibrium temperature as a function of $log(U)$ computed with TEPID and Cloudy (blue and red lines, respectively).}
\label{fig_UT}
\end{figure}

\begin{figure}
\centering
\includegraphics[width=\columnwidth]{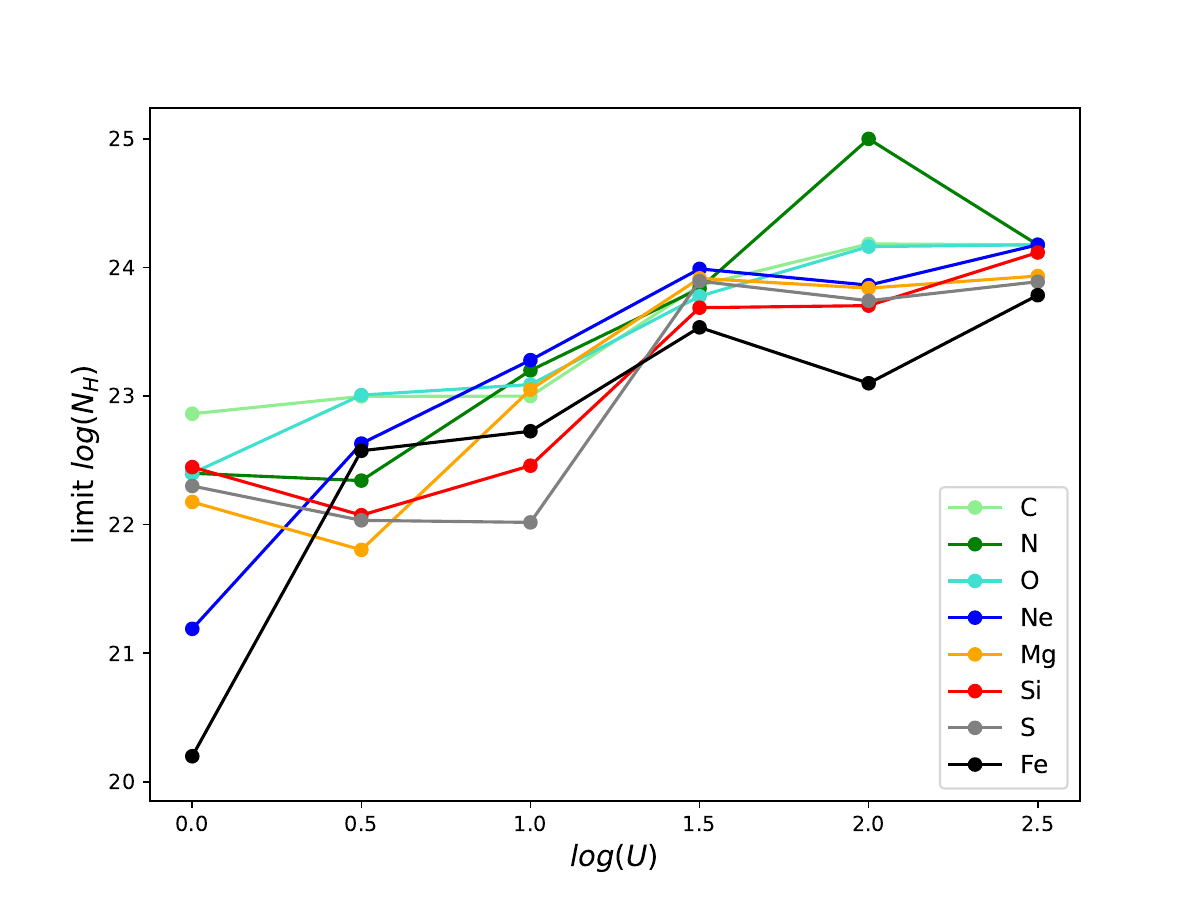}
\caption{Limit $N_H$ up to which TEPID and Cloudy predicted ionic columns are in agreement (i.e. within 50\% one another), as a function of $log(U)$.}
\label{fig_nions}
\end{figure}

Figure \ref{fig_UT} shows the equilibrium gas temperature as a function of $log(U)$ for the NGC 4051 SED (described in Sect. \ref{4051 sims} below), computed with both TEPID (blue) and Cloudy (red), with $log(N_H/cm^{-2})=18, log(n_H/cm^{-3})=6$. The two codes are in quite good agreement over six decades in $log(U)$.
We also compare the radiative transfer of TEPID and Cloudy. For a given $log(U)$, we run both codes for a total $log(N_H/cm^{-2})={25}$ and compute up to which depth the ionic columns of a given metal agree within 50\%. To have a meaningful comparison, we restrict to the most significant (and so detectable) ions, i.e. those with a fractional abundance $\geq 0.1$.
Fig. \ref{fig_nions} shows the result between $log(U)=0.0$ and 2.5. The agreement is overall quite good and the limit is above the typical space occupied by the ionised winds observed in AGNs. As discussed in \cite{luminari23b}, the discrepancies for S and Fe are driven by the lack of updated atomic data rather than intrinsic differencies between the two codes. To overcome this, Cloudy uses custom-computed means, while we employ the latest data available, i.e. \cite{m98} for Fe I-X and, starting from this version, \cite{2022A&A...668A..72B} for S I and Fe I.

\begin{figure}
\centering
\includegraphics[width=\columnwidth]{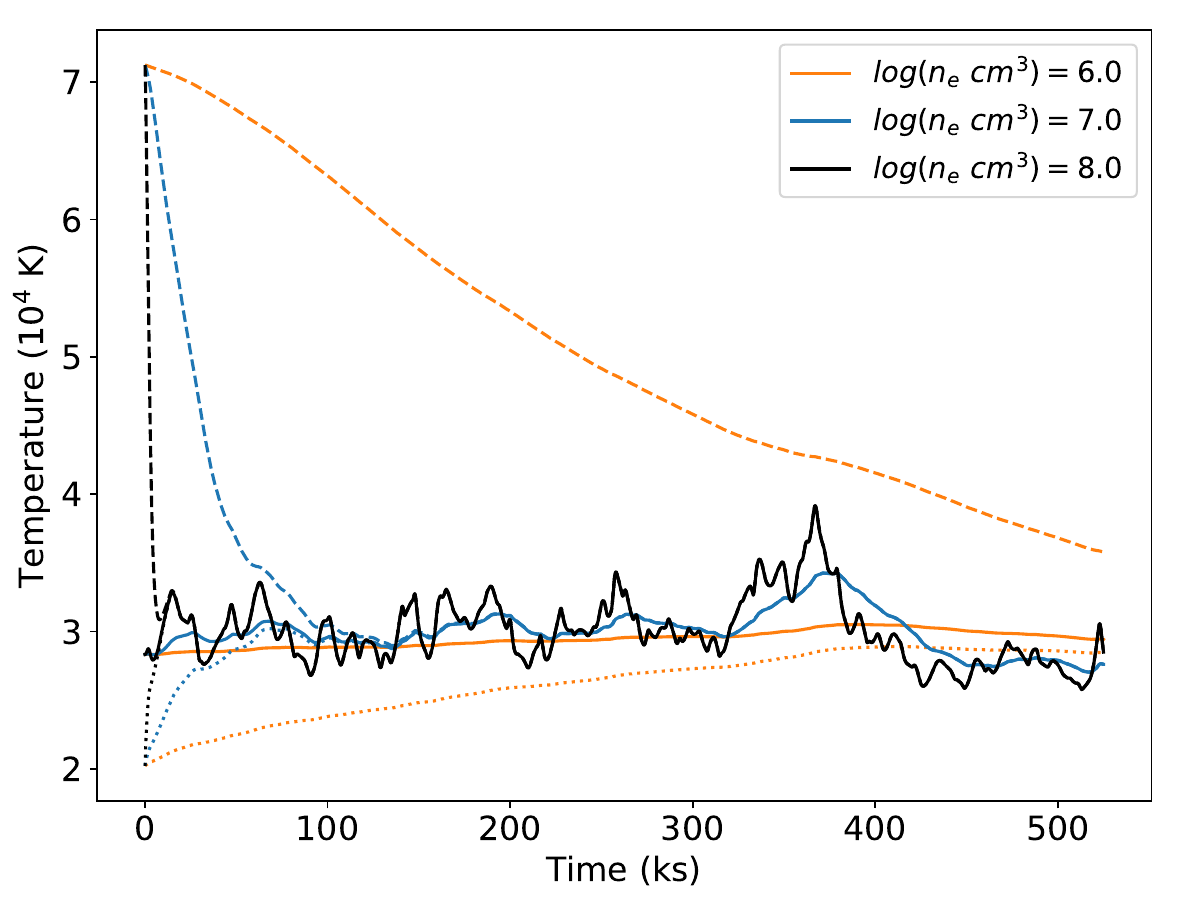}
\caption{Gas temperature as a function of time for $log(U_0)=0$ and the NGC 4051 lightcurve from Fig. \ref{fig:LC_XMM_Nustar}. Different colours correspond to increasing $log(n_e)$ (see legend). Solid lines are for $f_U=1$, while dotted(dashed) are for $f_U=0.2$(=5).}
\label{deltau}
\end{figure}

As TEPID is a first-order derivative integrator, it is necessary to specify the initial state of the gas at t=0, from which it is then evolved following the lightcurve. In the previous version of TEPID the gas was assumed to be in photoionisation equilibrium at the beginning of the temporal integration and, thus, both the ionisation and the photon field at t=0 were parametrised by the initial ionisation parameter $U_0$. It is important to stress that this did not strictly mean that the absorber was initially in photoionisation equilibrium, but rather that the ionic abundances were loosely clustered around those corresponding to a certain equilibrium value. For a given gas number density $n_e$, $U_0$ directly translated into a distance $r$ from the ionising source via Eq \ref{eq_r}. For $t>0$ the photon field was then varied according to the inputted lightcurve. 
To relax such assumption, TEPID now also includes an over/under-ionisation parameter $f_U$. The photon field at $t>0$ is still derived according to $U_0$ and then varied following to the lightcurve. However, the radiation at $t=0$ does not correspond to that for $U_0$, but to $f_U \cdot U_0$. Setting $f_U < 1$($> 1$) mimics, to a first order, an initial under-(over-) ionisation. 
Fig. \ref{deltau} shows an example from the NGC 4051 simulations detailed below. Solid lines show the temperature for $log(U_0)=0, f_U=1$ for $log(n_e /cm^{-3})=6,7,8$ (colour coded). At t=0 the temperature corresponds to the equilibrium one, T=$3 \cdot 10^4 K$, regardless of the number density, and then evolves according to the lightcurve and $log(n_e)$. Dotted(dashed) lines, instead, correspond to the same $log(n_e), log(U_0)$ but with $f_U=0.2$(=5). The initial temperatures correspond to those for $log(U_0')= log(f_U \cdot U_0) = -0.7(+0.7)$, i.e. T=$2\cdot 10^4 K$($7\cdot 10^4 K$). At later times, the temperature increases(decreases) and catches up that for $f_U =1$ with a lag inversely proportional to $n_e$, as expected.

\subsection{Simulations for NGC 4051}
\label{4051 sims}

We perform TEPID simulations over the whole NuSTAR lightcurve, setting as t=0 the start of the NuSTAR observation, for a fine grid of input values of $log(U_0), log(N_H), log(n_e), f_U$.
We span a range in ionisation $\log(U_0) \in [-1.5,2.5]$, with a step of 0.125. The density range is $\log(n_e/cm^{-3}) \in [5,10]$ with a step of 0.25, covering the whole dynamical range: the lower bound corresponds to a gas with no ionisation variation over the observed time, while the upper bound varies so quickly that is practically in ionisation equilibrium within each time-resolved spectra. Based on the maximum X-ray variability in Fig. \ref{fig:LC_XMM_Nustar}, we explore a range of under/over-ionisation within a factor of 4: $f_U = [1/4,1/2,1,2,4]$. The simulations are run from $log(N_H/cm^{-2})=20$ (as the gas is optically thin below this value) to 22, with steps of 0.05 to accurately resolve the radiative transfer. These values are tailored to the best-fit results of the time-averaged XMM-Newton RGS spectra from Paper I. 
The predicted time-resolved ionic columns will then be fed to the PHASE engine to be directly fitted to the X-ray spectrum of NGC 4051. This is the same exercise as usually done with ionisation equilibrium codes, but introducing $n_e$ as a free parameter, which dictates the variability timescale and, thus, the spectral variability of the absorbers across the time-resolved spectral bins. 

\begin{figure}
\centering
\includegraphics[width=\columnwidth]{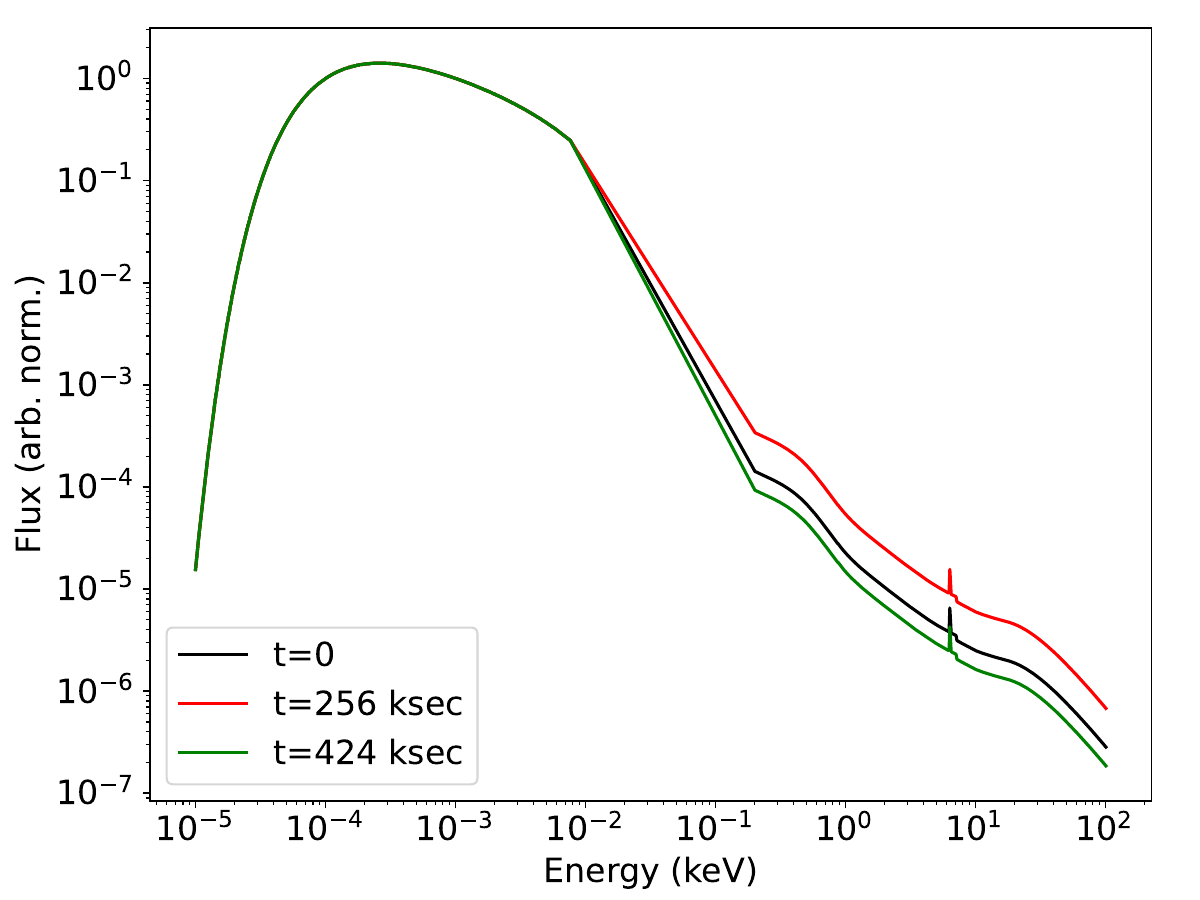}
\caption{TEPID input SED at three different times: black line is at t=0, while red and green lines are at the time of the highest and lowest XMM-Newton observed fluxes, respectively.}
\label{fig:sed}
\end{figure}

As we will discuss in Sect. \ref{sec_deltau}, starting the simulations at the beginning of NuSTAR pointing, 247 ksec before XMM, allows the gas to (partially) relax from the initial assumptions and start following the luminosity variations ahead of the time-resolved spectra (see Fig. \ref{deltau} and related discussion). 
We use the standard AGN SED from Cloudy until the XMM OM datapoint for Obs1 (at E=7 eV). Then, from 0.3 to 100 keV we adopt the best-fitting continuum to the time-averaged spectrum of Obs1 (i.e. a black body, a powerlaw and a cold reflector; see Appendix \ref{appendix_timeavg} for all the details), whose parameters (photon index, temperature and normalisations) are in excellent agreement with the median values that will be found for the time-resolved spectra
(Sect. \ref{Data Analysis} and Table \ref{tab_timeres}). The two portions of the SED, below 7 eV and above 0.3 keV, are connected with a powerlaw to ensure continuity.
The first portion is kept constant, since the OM data show negligible variation (only 2\% between Obs1 and Obs2), while the second portion is scaled rigidly according to the NuSTAR 3-6 keV lightcurve\footnote{Note that the 3-79 keV lightcurve is practically identical to the 3-6 keV one}. The powerlaw connecting the two portions is updated accordingly. Figure \ref{fig:sed} shows the SED at t=0 and at the highest and lowest fluxes in the XMM-Newton pointings, respectively at t=256 and t=424 ksec. 

It must be noted that similar results (and fully consistent best-fit solutions) would be obtained by scaling rigidly the whole SED. This is because the low energy part (i.e. until the OM datapoint) mainly contributes to the temperature balance (through the Compton heating-cooling term). For the limited temperature excursion given by the NGC 4051 lightcurve (see Fig. \ref{deltau} as a reference), the associated variations in the recombination rates are quite limited. On the other hand, the ionisation is mainly regulated by the high energy, powerlaw-like portion of the SED above E=0.3 kev.

\section{Data Analysis}
\label{Data Analysis}
\subsection{Time-resolved spectral fit}
\label{sect-fit}

As discussed above, Paper I identified the presence of three absorbers using the Obs1 and Obs2 time-averaged RGS spectra. We now build on their best-fit model to fit the 37 time-resolved intervals, each one including an EPIC-pn, a NuSTAR-FPMA and -FPMB spectrum. Their continuum model consisted of a black body and a powerlaw. We add a \texttt{xillver} component to account for the reflection, mainly constrained by NuSTAR at $E\gtrsim 10 keV$, and the Fe K$\alpha$ emission line. We also include the O VII, O VIII, Ne IX ($E \approx560,653,906$ eV) lines found in the RGS data, with energies, equivalent widths and broadenings frozen at their best-fit values. A Galactic absorber with column density fixed to $N_H=1.2 \cdot 10^{-20} cm^{-2}$ is adopted, as per \cite{HI4PI}.
The three absorbers are modelled with our customised version of the PHASE spectral enginge, fed with the TEPID time-resolved ionic columns. Besides the free parameters of TEPID, in PHASE there are two more parameters: the line-of-sight velocity $v_{out}$ and the line broadening $\sigma_v$. Each component has $log(U_0),\log(n_e/cm^{-3})$ linked between the 37 time-resolved intervals. We keep $N_H, v_{out}$ fixed to the RGS best-fit values, in analogy with the strategy of K07, and we set $\sigma_v=200 km\ s^{-1}$ to represent an unresolved line at the EPIC-pn and NuSTAR energy resolution. We fix the \texttt{xillver} inclination $i$ to a fiducial value of 30 $deg$.
The model can be expressed as:

\begin{eqnarray}
    galabs \cdot const \cdot (LIP \cdot HIP \cdot HVIP \cdot (bb+powerlaw \nonumber\\
    +xillver)+gauss_{OVII}+gauss_{OVIII}+gauss_{NeIX})
    \label{eq:model}
\end{eqnarray}

where \textit{const} accounts for the intercalibration between the different telescopes. It is set to 1 for EPIC-pn and it is linked between the FPMA and FPMB spectra. The redshift of the continuum components is fixed to the systemic one. Spectral analysis is performed with the \textit{sherpa} \citep{sherpa} fitting package with $\chi^2$ statistics. All errors hereafter are at 1 $\sigma$, unless stated.

\begin{table}[]
\centering
\begin{tabular}{l|c }
Parameter & Value \\
\hline
\hline
\textbf{galabs} \\
$N_H$ & $1.2\cdot 10^{20}$($^f$)\\
\textbf{const} (fpma) & $1.14 \pm 0.01$($^l$) \\
\textbf{const} (fpmb) & $1.14 \pm 0.01$($^l$) \\
\hline
\textbf{LIP} \\
$\log(U_0)$ & $-0.03^{+0.05}_{-0.06}$ \\
$\log(n/cm^{-3})$ & $<5.5$ \\
$\log(N_H/cm^{-2})$ & 20.2 ($^f$) \\
$v_{out}$ (km\ s$^{-1}$) & 400 ($^f$) \\
$r$ (pc) & $>2.5 \cdot 10^{-2}$ \\
$\dot{M}_{out}$ ($\dot{M}_{\odot}\ yr^{-1}$) & $>1.4\cdot10^{-4}$ \\
$\dot{M}_{out}/\dot{M}_{acc}$ & $>0.03$ \\
$\dot{E}_{out}$ ($erg\ s^{-1}$) & $>7.8\cdot10^{39}$ \\
\textbf{HIP} \\ 
$\log(U_0)$ & $1.57^{+0.01}_{-0.02}$ \\
$\log(n/cm^{-3})$ & $7.0^{+0.5}_{-0.2}$ \\
$\log(N_H/cm^{-2})$ & 21.3 ($^f$) \\
$v_{out}$ (km\ s$^{-1}$) & 530 ($^f$) \\
$r (pc)$ & $6.85^{+0.05}_{-0.02} \cdot 10^{-4}$ \\
$\dot{M}_{out}$ ($\dot{M}_{\odot}\ yr^{-1}$) & $3.02 \pm 0.01\cdot10^{-5}$ \\
$\dot{M}_{out}/\dot{M}_{acc}$ & $6.84 \pm 0.03 \cdot 10^{-3}$ \\
$\dot{E}_{out}$ ($erg\ s^{-1}$) & $1.71 \pm 0.01\cdot10^{39}$ \\
\textbf{HVIP} \\ 
$\log(U_0)$ & $2.37^{+0.07}_{-0.03}$ \\
$\log(n/cm^{-3})$ & $>8.7$ \\
$\log(N_H/cm^{-2})$ & 21.4 ($^f$) \\
$v_{out}$ (km\ s$^{-1}$) & 5800 ($^f$) \\
$r (pc)$ & $<4.0 \cdot 10^{-5}$ \\
$\dot{M}_{out}$ ($\dot{M}_{\odot}\ yr^{-1}$) & $<9.0\cdot10^{-6}$ \\
$\dot{M}_{out}/\dot{M}_{acc}$ & $<2 \cdot 10^{-3}$ \\
$\dot{E}_{out}$ ($erg\ s^{-1}$) & $<1.5\cdot10^{37}$ \\
\textbf{bb} \\
kT (eV) & $98.9 \pm 0.4$ \\
norm $^{a,m}$ & $1.26(1.33)\pm 0.05$\\
z & $2.34 \cdot 10^{-3}$($^f$) \\
\textbf{powerlaw} \\
$\Gamma\ ^m$ & $1.83(1.85) \pm 0.02$ \\
norm $^{b,m}$ & $4.12(4.56)\pm 0.08$ \\
\textbf{xillver} \\
$\Gamma$ (obs1) & $1.93 \pm 0.02$ \\
$\Gamma$ (obs2) & $1.81^{+0.04}_{-0.03}$\\
$A_{fe}$ & $0.73\pm0.05$($^l$) \\
$E_{cut}$ (keV) & $>640$($^l$) \\
$\log(\xi)$ & $<0.11$($^l$) \\
z & $2.34 \cdot 10^{-3}$($^f$) \\
i (deg) & 30 ($^f$) \\
norm (obs1) $^c$ & $1.25 \pm 0.03$ \\
norm (obs2) $^c$ & $0.67\pm0.03$ \\
\hline
\hline
$\chi^2/$ deg. of freedom \\
Total (obs1+obs2) & 10140.5/15633 (=0.65) \\
\end{tabular}
\caption{Best-fit values of the time-resolved EPIC-pn and NuSTAR fit. Values are always linked between the three instruments within a given observation, except for the normalisation constants. ($^f$): fixed value; ($^l$): value linked between obs1 and obs2; ($^m$): median(mean) value across the 37 observations with the median error (the mean error is always within 10\% of the median one). Normalisations are in units of: $^a$, $10^{-4} L_{39}/D_{10}$, where the ratio is between the luminosity in units of $10^{39} erg/s$ and the distance in units of 10 kpc; $^b$, $10^{-3}$ ph/keV/cm$^2$/s; $^c$, $10^{-4}$.}
\label{tab_timeres}
\end{table}

\begin{figure}
\centering
\includegraphics[width=\columnwidth]{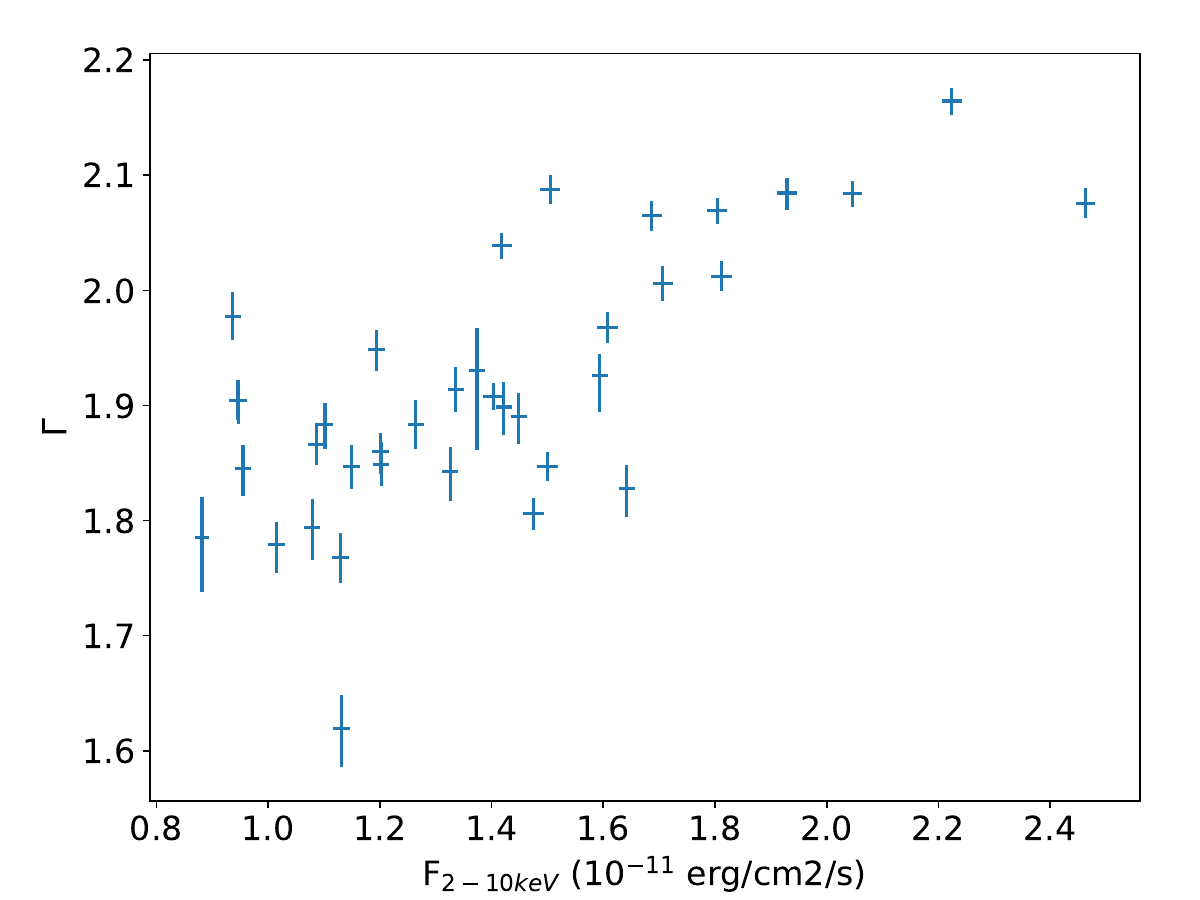}
\caption{Photon index $\Gamma$ of the primary powerlaw as a function of the 2-10 keV flux.}
\label{fig_gamma}
\end{figure}

\begin{figure}
\centering
\includegraphics[width=\columnwidth]{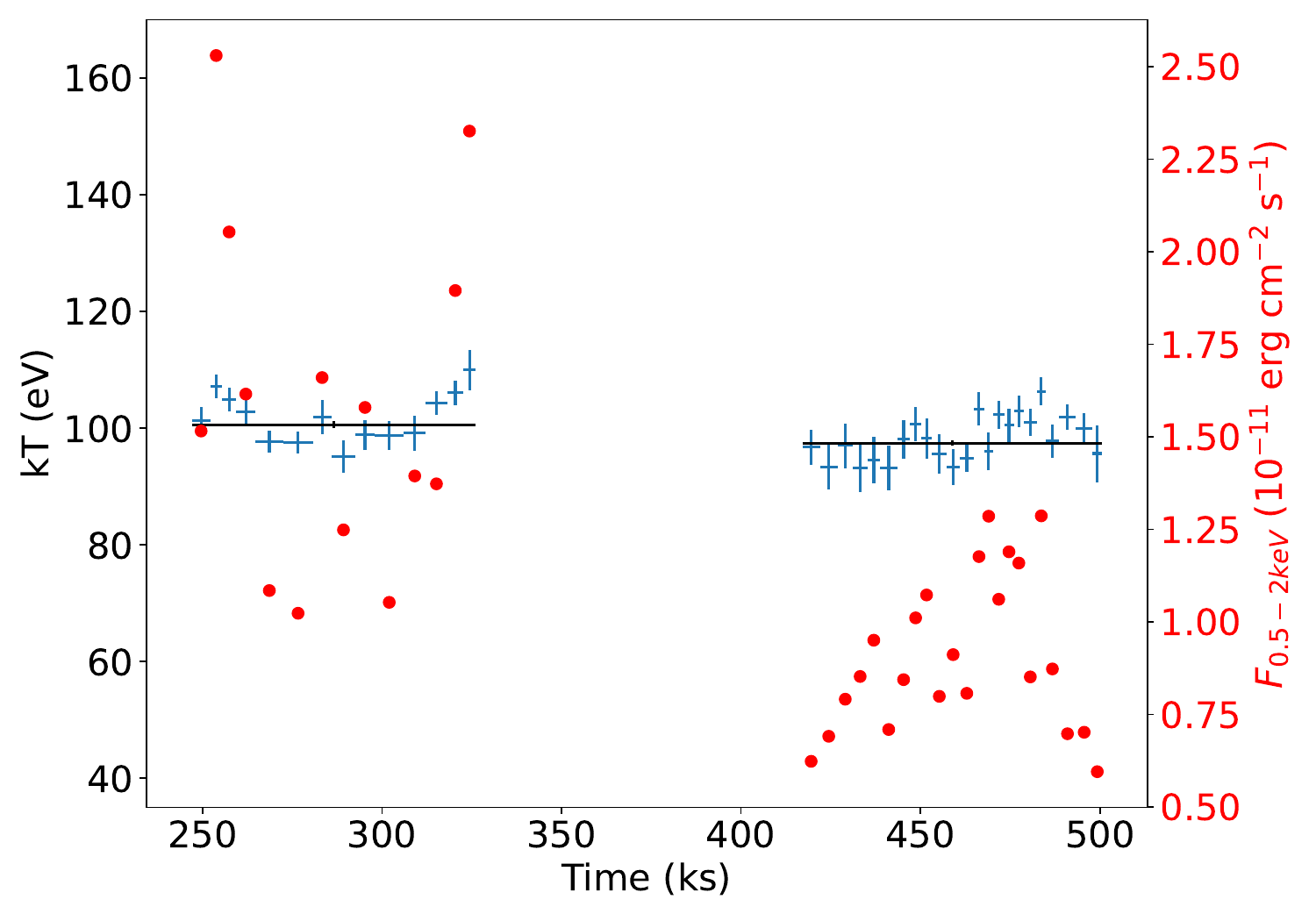} \\
\includegraphics[width=\columnwidth]{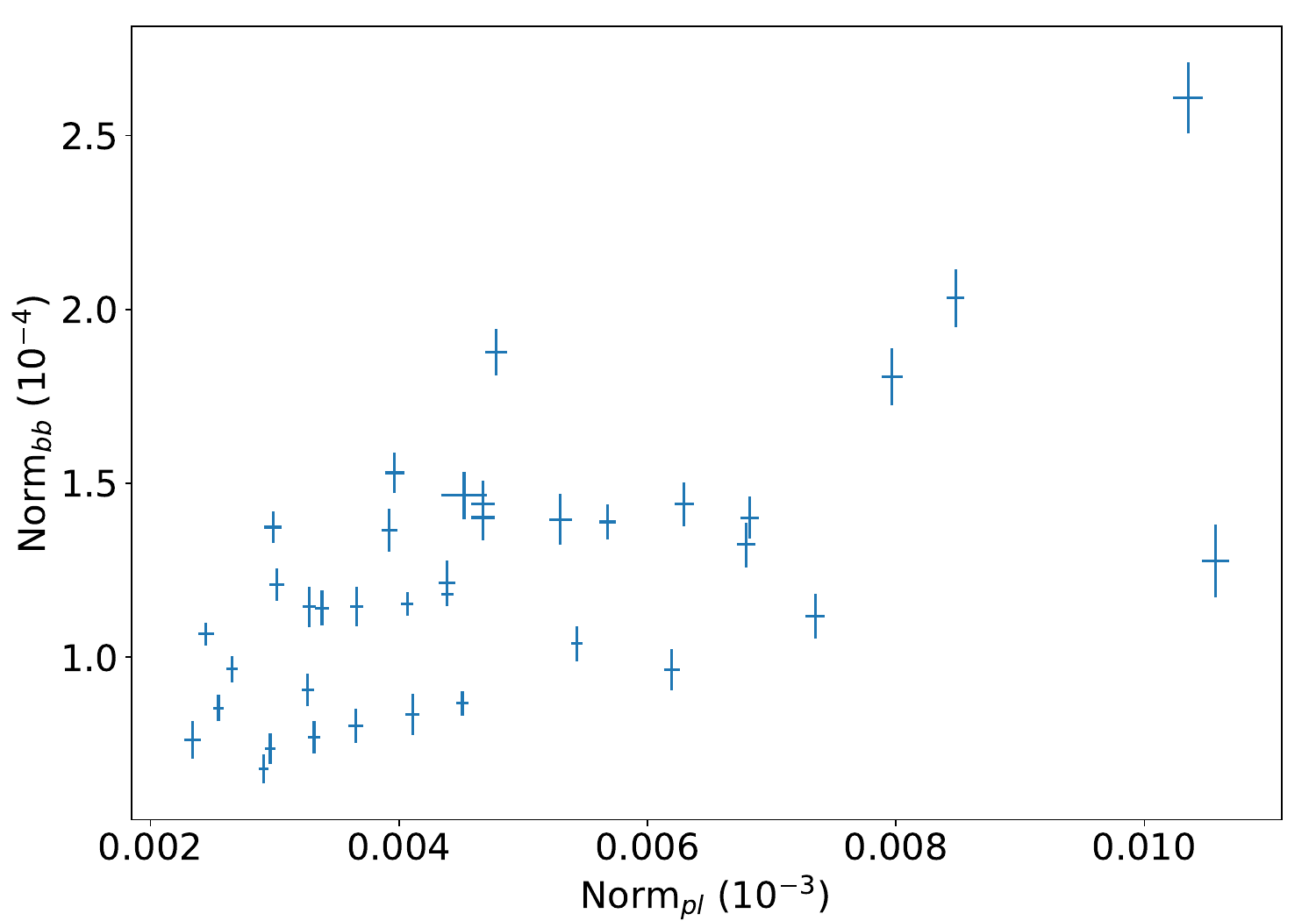}
\caption{Top: black body temperature as a function of the observing time (left y-axis). Red dots report the 0.5-2 keV luminosity in each temporal bin (right y-axis). Bottom: norm$_{bb}$ as a function of norm$_{pl}$.}
\label{fig_bb}
\end{figure}

To reduce parameter degeneracy, we first freeze all the absorbers at their RGS best-fit values and inspect the variability of the continuum by letting the powerlaw, black body and \texttt{xillver} free to vary for each time bin. 
As shown in Fig. \ref{fig_gamma}, the powerlaw exhibits a "softer when brighter" behaviour, i.e. $\Gamma$ positively correlates with $F_{2-10 keV}$. Such correlation is commonly observed in variable AGNs and it is thought to be due to a more efficient cooling of the X-ray reprocessing "corona" when the primary accretion disc radiation increases (e.g. \citealp{2009MNRAS.399.1597S,2017A&A...600A.101S}). Fig. \ref{fig_bb}, top, shows the blackbody temperature $kT$ as a function of the observing time (left y-axis). Blue points report the best-fit values for each time bin, while black points are the average for Obs1 and Obs2. As a comparison, red dots (right y-axis) show the variation of the 0.5-2 keV flux in each time bin. Bottom panel shows the black body normalisation $norm_{bb}$ as a function of the powerlaw normalisation $norm_{pl}$. While $kT$ is constant (within the errors), $norm_{bb}$ shows significant variations, quite tightly correlated with $norm_{pl}$. This finding is in line with earlier results for this source \citep{pounds04,uttley04}.

\begin{figure*}
\centering
\includegraphics[width=1.8\columnwidth]{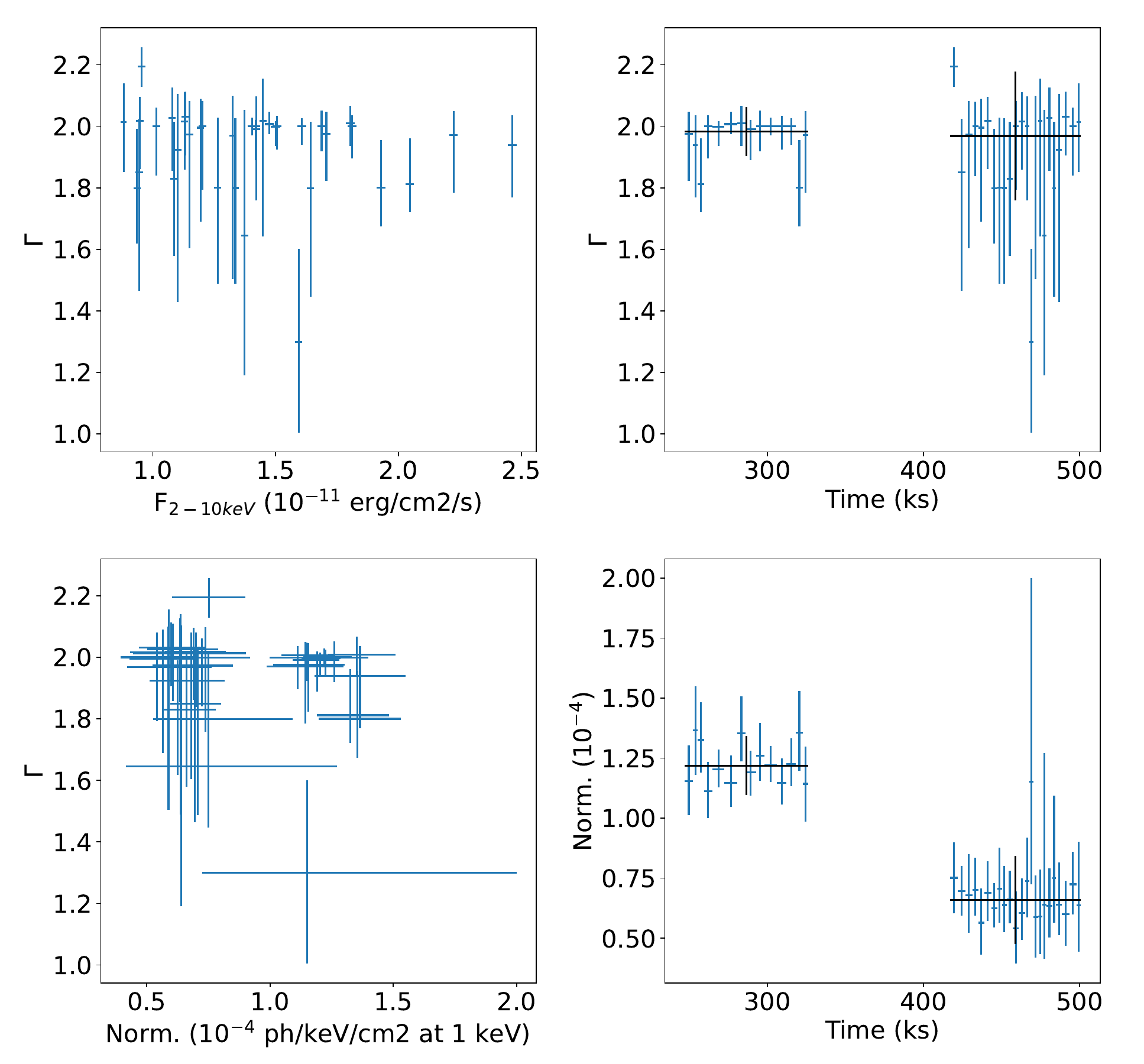}
\caption{Trend of $\Gamma$ and normalisation of the reflection component. Left column: $\Gamma$ against 2-10 keV flux (top) and normalisation (bottom). Right column: $\Gamma$ (top) and normalisation (bottom) as a function of the observing time.}
\label{fig_refl}
\end{figure*}

Regarding the reflection component, $\Gamma$ does not follow any immediate trend neither with $F_{2-10 keV}$ nor with its own normalisation (see Fig. \ref{fig_refl}, top and bottom left panels, respectively) and it is consistent with being constant throughout the observing time (top right panel). More interestingly, the normalisation is constant in Obs1 and Obs2 but shows a significant variation between them. 
This indicates that the reflector is located at distances $>8 \cdot 10^4$ light-second (the duration of each of the two observations) and $< 1.75 \cdot 10^5$ light-second (the separation between the mid-points of the two observations), i.e. 8500–18600 $r_G$. For NGC 4051, the gravitational radius is $r_G=G M_{BH} /c ^2 = 2.08 \cdot 10^{11} cm = 6.73 \cdot 10^{-8} pc$, where we assume a black hole mass $M_{BH}=1.4 \cdot 10^6 M_{\odot}$ as in Paper I (from \citealp{2000ApJ...542..161P}, hereafter P00). 
The reflecting material is slightly underabundant in Fe ($A_{fe}=0.73$; but please see \citealp{ding24} for the impact of high gas density on this parameter) and cold, with a best-fit ionisation parameter consistent with the lower tabulated bound of $\log(\xi/erg\ cm\ s)=0$. The energy cutoff $E_{cut}$ cannot be constrained by the present data.

\begin{figure*}[]
\centering
\includegraphics[width=12.0cm]{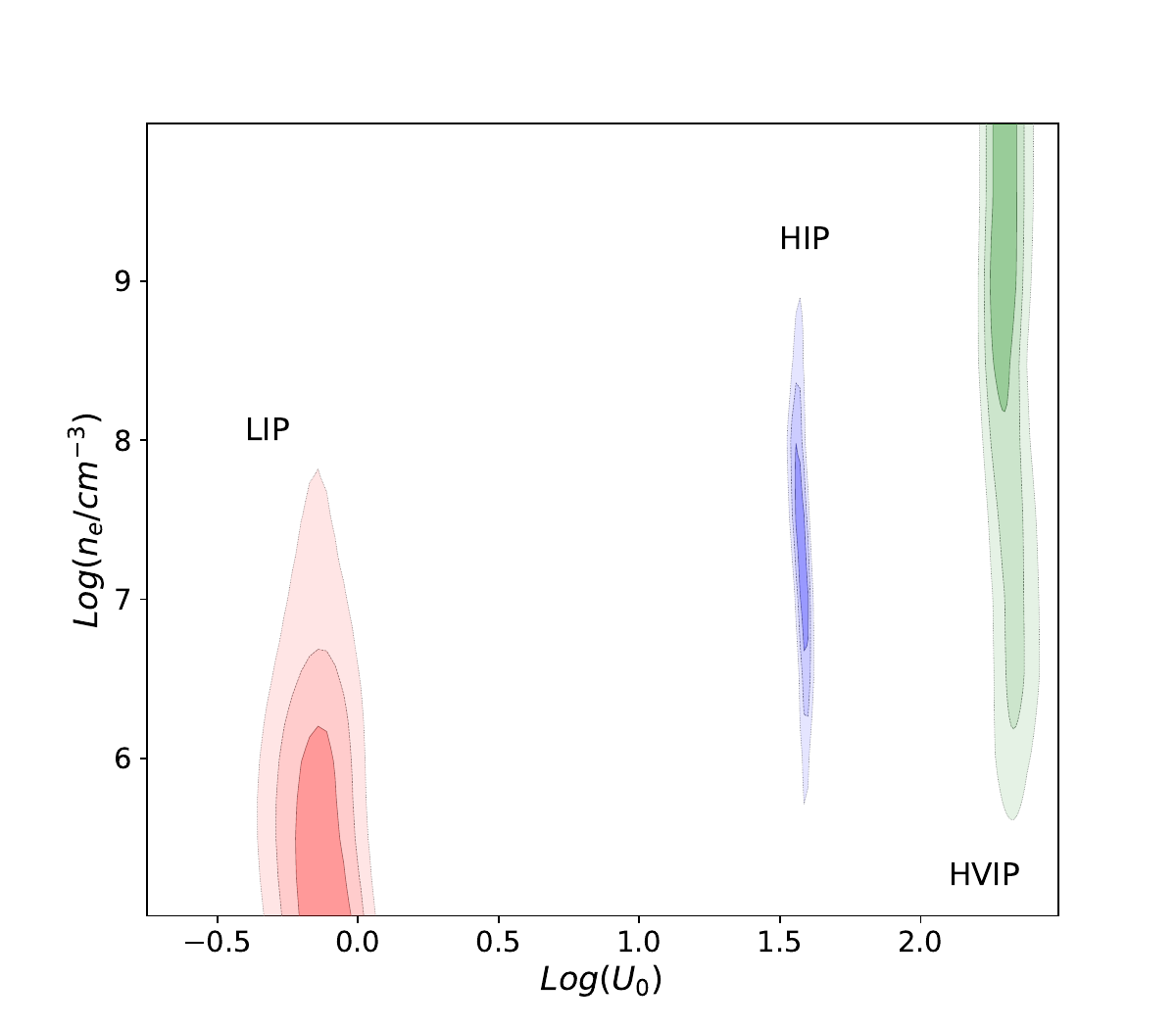} 
\caption{Contour plot of the ionisation $Log(U_0)$ versus the number density $log(n_e/cm^{-3})$ (x- and y-axis, respectively) for the LIP, HIP and HVIP modelled with TEPID. The levels correspond to 1,2,3 $\sigma$ significance.} 
\label{fig:Contours_3absorbers_TEPID}
\end{figure*}

Based on these results, we then leave $\Gamma$ and $norm_{pl}$ of the primary powerlaw free to vary in each time bin, while the reflection parameters are linked between the first 14 time bins (corresponding to Obs1) and the last 23 (Obs2). The black body $kT$ is linked among all the time bins, while $norm_{bb}$ is free to vary independently in each time bin. After performing a first fit of these continuum components, we leave $log(U_0), f_U, log(n_e)$ free for LIP, HIP, HVIP and fit again. $f_U$ is linked between the three WAs, as they all "see" the same incident continuum. The $log(U_0)-log(n_e/cm^{-3})$ contour plot for the three components is shown in Fig. \ref{fig:Contours_3absorbers_TEPID}, with levels corresponding to 1,2,3 $\sigma$ significance. We obtain upper and lower limits on the number density for the LIP and HVIP components, $log(n_e/cm^{-3})<5.5$ and $>8.7$. The HIP is fully constrained, $log(n_e/cm^{-3})=7.0^{+0.5}_{-0.2}$, and in agreement with what found in Paper I (albeit with smaller errors). 
Table \ref{tab_timeres} reports the best-fit values. 
The derived values of $r$ and of the energetic are discussed in the following sections. For the sake of simplicity, we only report the mean and median values of the primary $\Gamma$, $norm_{pl}$ and of $norm_{bb}$ across the 37 time bins.

\subsection{Time evolution of the opacity and ionisation of the HIP component}
\begin{figure}
\centering
\includegraphics[width=\columnwidth]{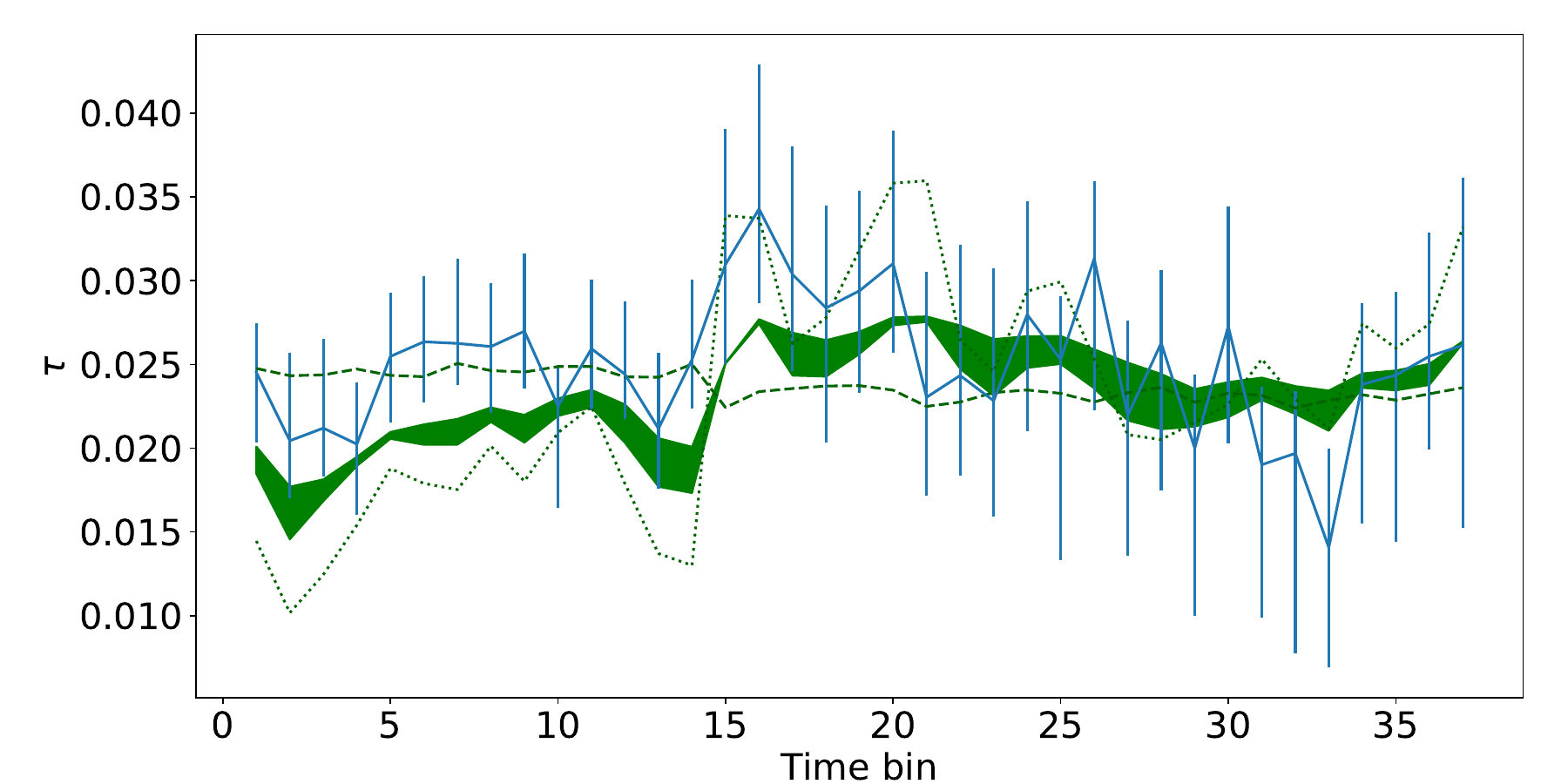}
\caption{Opacity lightcurve for the HIP. The shaded green area corresopnds to the $\pm 1 \sigma$ best-fit confidence interval for $log(n_e)$. Dashed and dotted lines are for $log(n_e/cm^{-3})=5,10$, respectively. Blue points and errorbars, instead, are opacity derived by fitting the spectra with time-equilibrium photoionisation models.}
\label{opaplot_HIP}
\end{figure}

\begin{figure}
\centering
\includegraphics[width=\columnwidth]{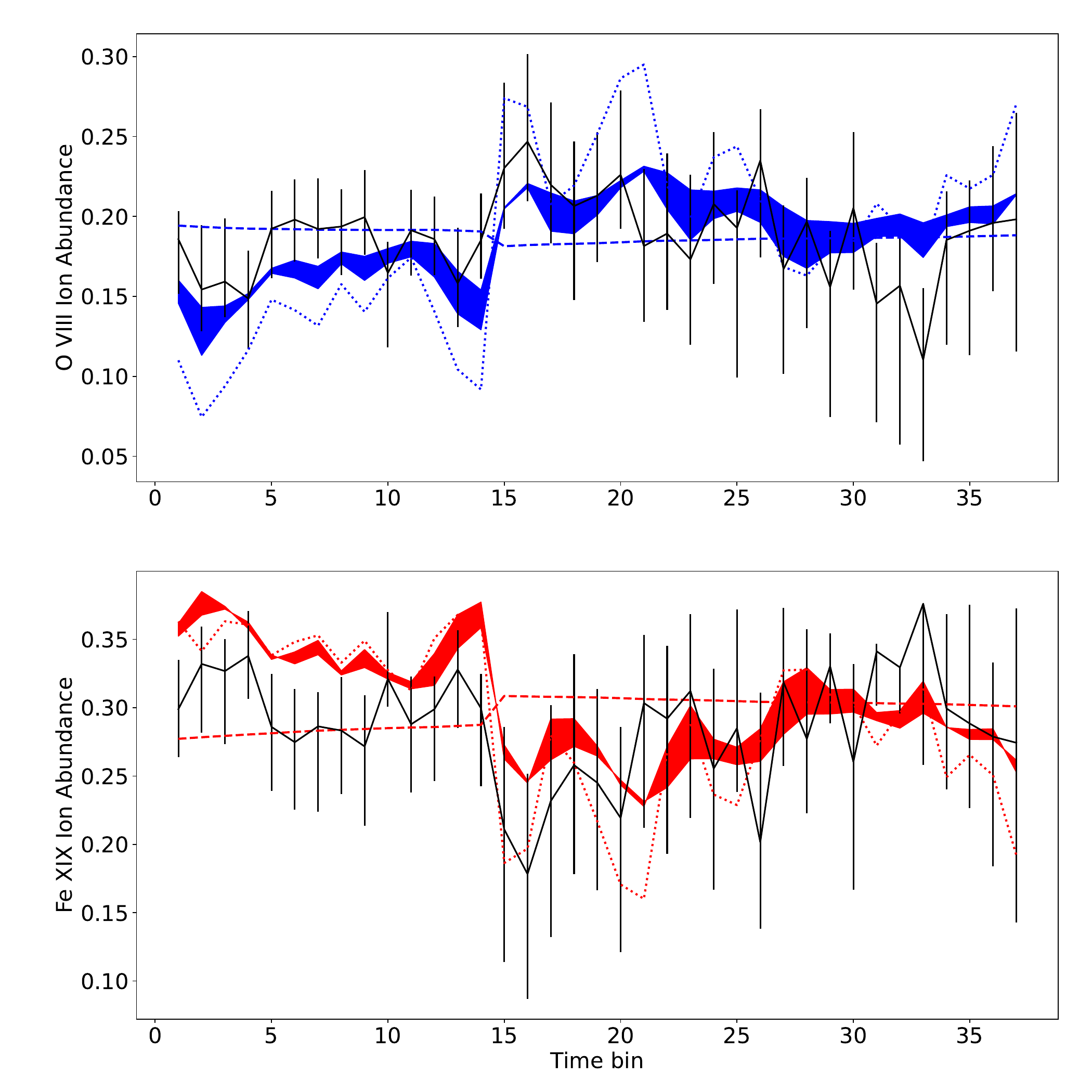}
\caption{Same as Fig. \ref{opaplot_HIP} but for the ionic abundances for O VIII (top) and Fe XIX (bottom). Blue and red bands correspond to the $\pm 1 \sigma$ best fit intervals for $log(n_e)$, while dashed and dotted lines are for $log(n_e/cm^{-3})=5,10$, respectively. Black points and errorbars are the best-fit and the associated $\pm 1\sigma$ uncertainties obtained with equilibrium models.}
\label{abunds_HIP}
\end{figure}

To better understand the impact of the time evolving ionisation on the observed spectra, we focus on the HIP component, which has a fully constrained $log(n_e)$. We plot in Fig. \ref{opaplot_HIP} the $\pm 1 \sigma$ interval in $log(n_e)$ of its energy-averaged opacity $\tau$ as a function of time (green area). $\tau$ is computed as:
\begin{equation}
\tau=-\ln \Big( \frac{ \int_{0.5}^2 F_{abs}\ dE}{\int_{0.5}^2 F_{cont}\ dE} \Big)
\end{equation}
where $F_{abs}, F_{cont}$ are the best-fit models including and excluding LIP, HIP, HVIP, respectively. The integration is carried out between 0.5 and 2 keV, where the absorption is significant (see Paper I and Appendix \ref{appendix_timeavg}). Green dashed and dotted lines report $\tau$ for the lowest and highest tabulated number density, $log(n_e/cm^{-3})=5$ and 10 respectively. For comparison, blue points correspond to the best-fit opacity obtained by replacing the time-evolving HIP component with 37 photoionisation equilibrium components (one per each spectrum), still computed with TEPID. This is just meant to obtain a measure as closest as possible of the "instantaneous" $\tau$ observed in each time bin.
The $\pm 1 \sigma$ green shaded region is the one that minimises the distance (from a statistical point of view) with the best-fitting equilibrium opacities of each spectral bin. Both the lowest and the highest $n_e$ are either too slow or to fast, thus resulting in a more statistically distant $\tau$.
Analogous trends are observed in the temporal evolution of the ionic abundances. Fig. \ref{abunds_HIP} shows the temporal evolution of two of the ions yielding the strongest absorption lines for the HIP (see Paper I), i.e. O VIII and Fe XIX. As above, points are for the equilibrium fit and coloured bands (and dashed lines) are for the time-evolving fit. Again, the time-evolving best-fit solution is the one that minimises the distance from the "instantaneous" abundances computed with the equilibrium components.

\subsection{Over/under-ionisation of the gas at t=0}
\label{sec_deltau}
\begin{figure}
\centering
\includegraphics[width=\columnwidth]{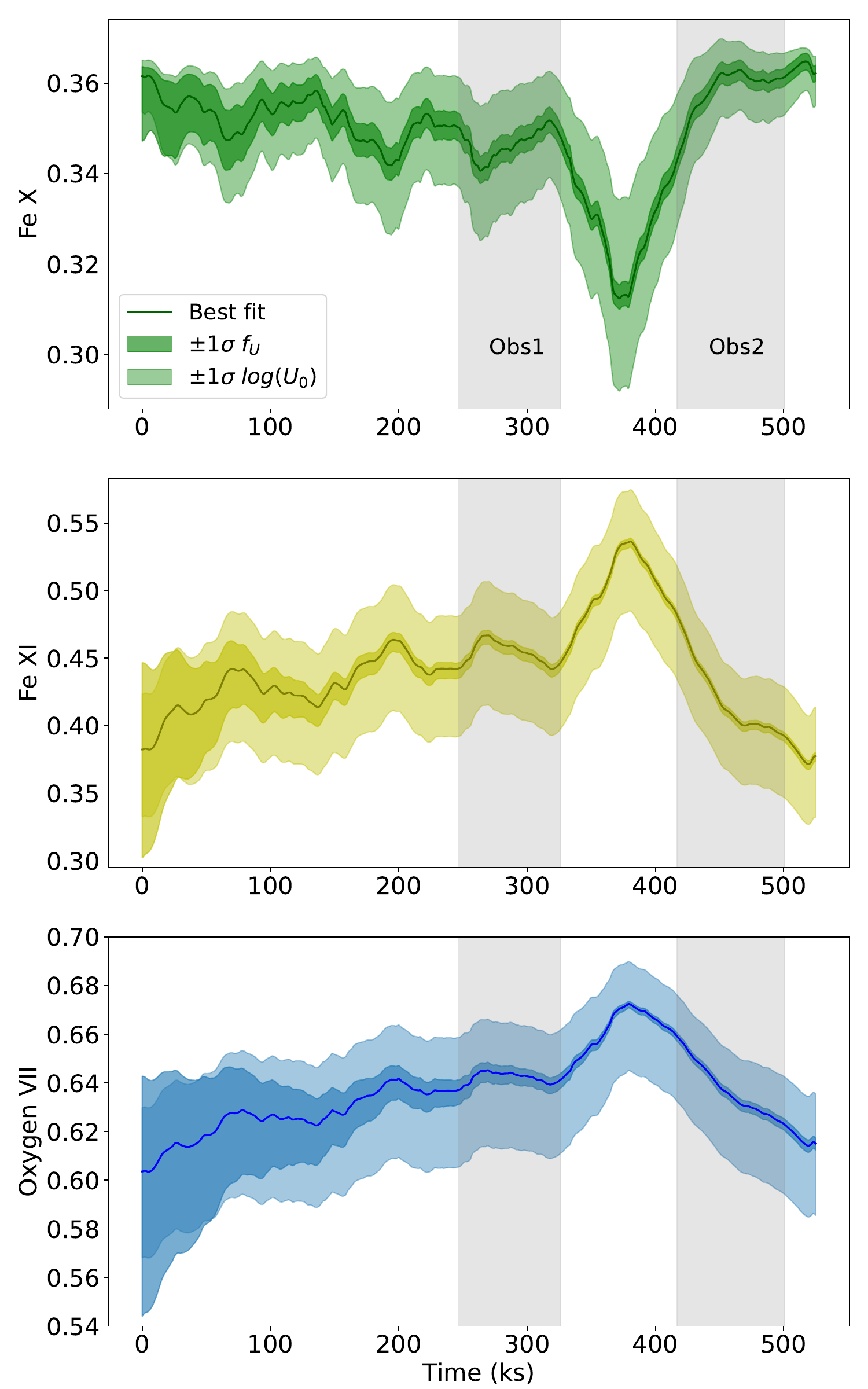}
\caption{Fe X, Fe XI, O VII ion abundances (top to bottom) for the LIP component, as a function of the observing time. Solid line is the tabulated solution closest to the best-fit values, while dark- and light-shaded regions are the $\pm 1 \sigma$ interval for $f_U$ and $logU_0$ respectively. Grey vertical intervals correspond to Obs1 and Obs2.}
\label{deltau_LIP}
\end{figure}

Leaving $f_U$ free to vary yields a best-fit value of $1.0 \pm 0.2$, with negligible variation of the other parameters. This can be easily explained by looking at the temporal evolution of the most significant ions. As anticipated in Fig. \ref{deltau}, the ionic distribution and the temperature tend to those for $f_U=1$ (i.e. initial equilibrium) for $t>0$. 
We explore in detail the effect of $f_U$ on the LIP, which has the lowest $n_e$ among the three components and so the longest $t_{eq}$.
Fig. \ref{deltau_LIP} shows the abundances of the ions providing most of the opacity for this component, i.e. Fe X, Fe XI and O VII, as a function of time (top to bottom). For each ion the solid line corresponds to the best fit solution, i.e. $log(U_0)=-0.03, log(n_e/cm^{-3})=5.5, f_U=1$. Dark dashed intervals correspond to the $\pm 1 \sigma$ interval for $f_U$, i.e. $\pm 0.2$, while light-shaded intervals correspond to the $\pm 1 \sigma$ interval for $\log(U_0)$. The vertical shadowed bands correspond to Obs1 and Obs2. For all the three ions, the interval associated to $f_U$ is well within that for $log(U)$ well before the start of Obs1. This means that an eventual over/under-ionisation at t=0 would lead to a variation of the ionic abundances significantly smaller than that associated to the best-fit uncertainty for $log(U)$.

\begin{figure*}
\centering
\includegraphics[width=1.4\columnwidth]{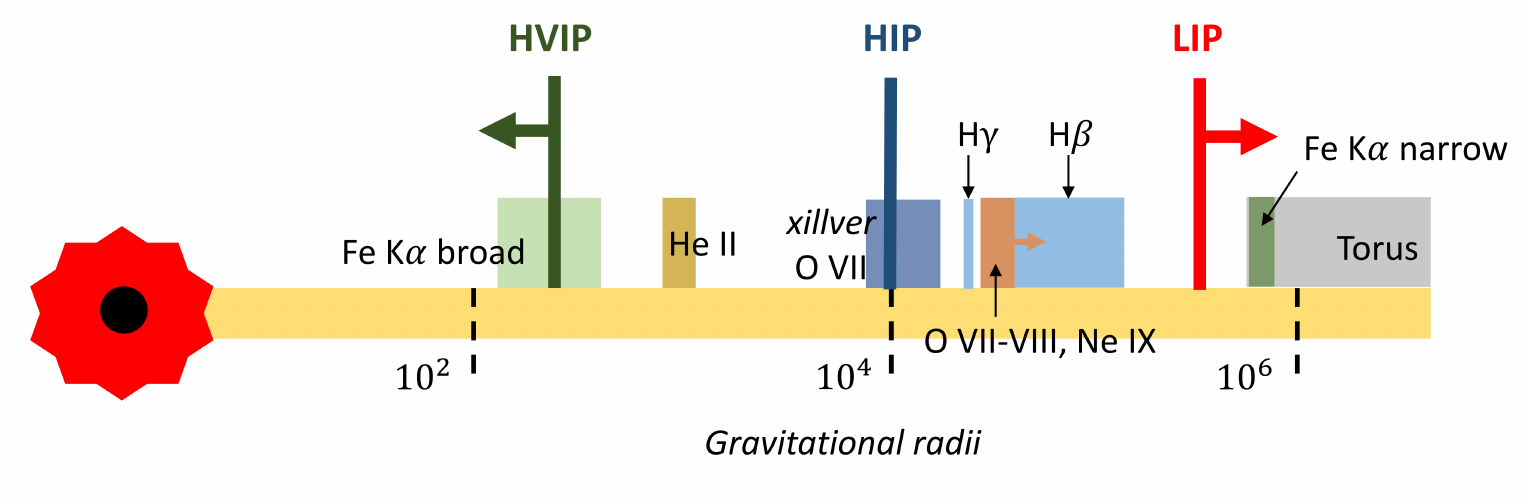}
\caption{Cartoon showing the relative radial locations (not to scale) of all the absorption and emission components in the innermost accretion disc region (see Sect. \ref{sec_radii}.)}
\label{fig:cartoon}
\end{figure*}

\section{Discussion \label{sec:Discussion}}
\subsection{Radial location of the absorption and emission components}
\label{sec_radii}
Thanks to the time-evolving fit we have been able to directly constrain $n_e$ for the HIP and to get upper(lower) limits for the LIP(HVIP). The distance $r$ between the inner face of the absorbers and the ionising source can be derived via Eq. \ref{eq_r}. HVIP, HIP, LIP distances are $\leq 600, (10.1^{+0.8}_{-0.3}) \cdot 10^4, \geq 3.8 \cdot 10^5 r_G$ (see upper part of Table \ref{tab_radii}). The radial thickness of the clouds can be derived as $\Delta r=N_H/n_H$. For all the components the relative thickness is very low, $\Delta r/r \leq 0.1$. 
Lower part of the Table reports the location of several components from optical, UV and X-ray observations from the literature discussed below. The cartoon in Fig. \ref{fig:cartoon} shows a scale diagram of all the components reported in the Table. For the sake of clarity, we divide the discussion in three spatial intervals:
\begin{itemize}
\item \textbf{Innermost accretion disc:} \cite{2011MNRAS.414.1965L} analysed a Chandra HETG \citep{2005PASP..117.1144C} 2008 spectrum and best-fitted the Fe K$\alpha$ line with a combination of a cold, unresolved component (EW=50 eV) together with a broad, FWHM=$16000^{+7000}_{-4000}$ km s$^{-1}$ component (EW=156 eV). Ascribing the broadening to rotational motion, a radial distance can be estimated as $r=G M_{BH}/FWHM^2=2.3 \cdot 10^{-5} pc$. Such results agree with those in \cite{2010ApJS..187..581S} from a 2000 \textit{Chandra} HEG spectrum. However, they fitted the line with just one component, finding indeed an EW consistent with the sum of the narrow and broad component and an intermediate broadening.
More recently, the high-resolution XRISM/Resolve \citep{tashiro20} spectrum analysed in \cite{reeves26} shows a Fe K$\alpha$ profile consistent with the above results. The broader component has FWHM=$11000^{+5000}_{-3300}$ km s$^{-1}$, implying $r=5.0 \cdot 10^{-5} pc$.
\item \textbf{Broad Line Region:} The optical reverberation campaign in P00 found broad He II $\lambda 4686$ and H$\beta$ emission lines with FWHM=5400$\pm$510 and 1100$\pm$190 km s$^{-1}$, respectively, corresponding to $r=2 \cdot 10^{-4}$ and $=2 \cdot 10^{-3}$ pc. Interestingly, He II centroid is blueshited by 1400 km s$^{-1}$ and with a prominent blue tail (width $\sim$ 2000 km s$^{-1}$). This led the authors to suggest a two-component model for the BLR: while the narrower and symmetric H$\beta$ line arises from a "flatlike structure" anchored to the accretion disk, He II is likely tracing an outflow, which gives rise to a blue tail and an overall blueshift due to projection effects. Subsequent work from \cite{2017ApJ...840...97F} found similar results; they also detected H$\gamma$ emission line from a distance slightly lower than that of H$\beta$. In both works, the distance of the H lines is comparable when estimated via the time lags. UV data from the Hubble Space Telescope STIS and COS instruments \citep{2001ApJ...557....2C,2012ApJ...751...84K} also show He II $\lambda$ 1640, C IV, N V, Si IV broad emission lines, best-fitted by a combination of two broad profiles (FWHM=1090 and 4500 km s$^{-1}$, similar to what found in the optical) and a narrower one (FWHM=260 km s$^{-1}$, which the authors associate to the Narrow Line Region). Several absorption lines due to C IV, N V, Si IV are also detected with outflowing velocities between 50 and 650 km s$^{-1}$ and have comparable ionisation and velocity to the X-ray absorbers.

Light-travel time allows to bracket $(0.8 \leq r \leq 1.7) \cdot 10^{-3} pc$ for the \texttt{xillver} X-ray reflector in Sect. \ref{sect-fit}. Concerning the soft X-ray lines, the non-variability of the O VII-VIII and Ne IX emission detected in the RGS in Paper I sets a lower limit for their distance of $2.4 \cdot 10^{-3} pc$. \cite{2019ApJ...879..102P} obtained a similar distance through the relative line intensities of the O VII emission complex in a 2009 low-state, high-signal to noise RGS spectrum, with associated $log(n_e/cm^{-3}) \approx 10$. The Ly$\alpha$ O VIII line in the same spectrum has FWHM $\sim$ 8900 km s$^{-1}$ and shows a remarkably similar profile (in velocity space) to the He II one. 
\item \textbf{Molecular torus:} The narrow component of the Fe K$\alpha$ in \cite{reeves26} and the associated K$\beta$ line have FWHM=$320^{+250}_{-200}$ km s$^{-1}$, corresponding to $r=0.06 pc$. This is coincident with the dust sublimation radius of NGC 4051: following \cite{2014MNRAS.443.1788C} and using a bolometric luminosity $L_{bol}=2.5 \cdot 10^{43} erg s^{-1}$ \citep{2004ApJ...606..151O} yields $r \approx 0.06 pc$.
\end{itemize}
All the present and archival observations disfavour a scenario in which HVIP and HIP are due to single outflowing clouds. As discussed in Paper I, HVIP has been detected as early as 2002 in Chandra LETG spectra \citep{2009A&A...496..107S}, as well as in a 2008 Suzaku observation \citep{2011MNRAS.414.1965L} with $v_{out},\ logU,\ N_H$ quite consistent with the present work. If the HVIP was due to a single cloud, it would have travelled a distance of 0.05 parsec per year, i.e. 0.8 pc until our 2018 observations. This distance is much larger than its best-fit $r< 4.0 \cdot 10^{-5} pc$ as well as the HIP location, $r= 6.8 \cdot 10^{-4}$ pc. On the other hand, the HIP is present at least since the 2001 observation analysed by K07 with similar properties and a consistent $r$ with what found here. In case of discrete outflowing clouds, observing such thin ($\Delta r/ r \leq 0.1$) geometries with such constant properties over more than one decade would require a fine tuning. Moreover, all components have both HVIP and HIP have $v_{out}$ much lower than the escape velocity $v_{esc}=\sqrt{2GM_{\odot}/r}$ at their locations, i.e. 5800 vs. 17400 km s$^{-1}$ for the HVIP, 530 vs. 4200 km s$^{-1}$ for the HIP and 400 vs 7000 km s$^{-1}$ for the LIP \footnote{For HVIP(LIP) we compute $v_{esc}$ corresponding to their upper(lower) limit on $r$}. It is worth noting that the velocities of the UV absorbers are also constant in the HST STIS and COS spectra, observed respectively in 1998 and 2009. 

Instead, the overall results are more suggestive of continuous flows with launching radii
clearly linked with the the emission components. First, the HVIP is co-spatial with the broad Fe K$\alpha$ component. Second, the X-ray reflection and the optical, UV and soft X-ray lines suggest a stratified Broad Line Region (BLR) extending roughly from $2 \cdot 10^{-4}$ to $2-5 \cdot 10^{-3}$ pc. \footnote{It must be noted that some of these lines (like He II $\lambda 4686$ and OVIII) are likely tracing a not fully virialised component, since they have a blueshifted centroid and a blue tail. As a result, their FWHM-derived locations may not be fully consistent; however, we use them as order-of-magnitude distance estimates.} (see \citealp{2025ApJ...995..200B,2024ApJ...973L..25X,2025ApJ...994L..13K} for stratification in other AGNs observed with XRISM). The HIP is located around the inner boundary of such BLR. Its best-fit $log(n_e/cm^{-3})=7.0$ is lower than what derived by \cite{2019ApJ...879..102P} for the O VIII and the reference value for BLR \citep{of06}, i.e. $log(n_e/cm^{-3})\approx 10.0$, again in agreement with a stratified stream, anchored to the (rotating) accretion disc and with an upper outflowing section whose density may decrease with the height. This spatial scale is the most constrained thanks to the optical data and the well-determined density and radius of the HIP. Finally, the narrow Fe K$\alpha$ component (from \citealp{reeves26}) and the dust sublimation radius, usually considered as the lower limit of the inner boundary of the torus, are both at $6 \cdot 10^{-2} pc$, close to the lower limit for the LIP, $r\geq 2.5 \cdot 10^{-2} pc$.

\subsection{Outflow energetics}
\label{sec_energetics}
For geometrically thin ($\Delta r/r << 1$) outflows the mass outflow rate can be computed as in K07:
\begin{equation}
\dot {M}_{out}=0.8 \pi m_p N_H v_{out} r f(\theta)
\end{equation}
where $f(\theta)$ is a geometrical factor to account for the projection along the line of sight and $m_p$ is the proton mass. We assume $f(\theta)=1.5$, as expected for typical Seyfert galaxies seen almost face-on (see discussion in K07 for more details). Values for LIP, HIP, HVIP are reported in Table \ref{tab_timeres}, both in units of $\dot{M}_{\odot}\ yr^{-1}$ and of the accretion rate $\dot{M}_{acc}=L_{bol}/\eta c^2$, where $\eta$ is the accretion efficiency (set =0.1, see e.g. \citealp{ss73}). Similarly to what found earlier for this source (K07), the mass rate is relatively modest and with a minor impact on the accretion rate. Energy outflow rate can be derived as $\dot {E}_{out}= (\gamma-1) \dot{M}_{out} c^2$ (where $\gamma=1/\sqrt{1-(\frac{v}{c}})^2$; see \citealp{llt21}). LIP, HIP, HVIP have energy rates around $10^{-6}-10^{-4} L_{bol}$ (see Table \ref{tab_timeres}), dramatically lower than the theoretical threshold of 0.5 - 5\% $L_{Edd}$ for them to have a significant impact on the host galaxy environment (\citealp{he10,Faucher12,King15}; $L_{Edd}$ is the Eddington luminosity and, for NGC 4051, $L_{Bol}=0.14\ L_{Edd}$). However, they may still be relevant in heating and chemically enriching the ISM, as discussed in K07. 
We also note that all WAs are located well inside the "escape radius" $r_{esc}$ at which their $v_{out}$ would correspond to the escape velocity (i.e, they have $v_{out} << v_{esc}$). Therefore, their are bound to become "failed winds" \citep{2019A&A...630A..94G,lne21,2026arXiv260416148M} and to fall back on the accretion disc. Otherwise, if the line-of-sight projected $v_{out}$ are much smaller than the real velocity $v_{tot}$, such that $ v_{tot}> v_{esc}$, the outflows can escape the gravitational well. This would imply highly inclined, almost vertical winds reminiscent of the "funnel-shaped" conical outflows as in \cite{elvis00}.
We also note that adopting $r_{esc}$ in the derivation of $\dot {M}_{out}$ and $\dot {E}_{out}$, as commonly done when no direct distance measurements are available, would lead to an overestimation of the energetics of the HVIP and HIP by factors of 1.7 and 2.8, respectively.

 \begin{table}[]
\centering
\begin{tabular}{l|c  c}
Component & $r$ (pc) & $r$  ($r_G$) \\
\hline
\hline
HVIP & $\leq 4.0 \cdot 10^{-5}$ & $\leq 600$ \\
HIP & $6.85^{+0.05}_{-0.02} \cdot 10^{-4}$ & $10.1^{+0.8}_{-0.3} \cdot 10^3$ \\
LIP & $\geq 2.5 \cdot 10^{-2}$ & $\geq 3.8 \cdot 10^5$ \\
\hline
Fe K$\alpha$ broad ($a,b$)&  $2.3-5.0 \cdot 10^{-5}$ & 3.4 - 7.4 $\cdot10^2$ \\
He II $\lambda 4686$ ($c1,c2$) & $2.0$ - $2.3 \cdot10^{-4}$ & 3.0 - 3.4 $\cdot10^3$ \\
\texttt{xillver} & $0.8 - 1.7 \cdot 10^{-3}$ & $1.1 - 2.5 \cdot 10^4$ \\
O VII ($d$) & $0.7 - 1.6 \cdot 10^{-3}$ & $1.0 - 2.4 \cdot 10^4$ \\
O VII-VIII, Ne IX ($e$) & $\geq  2.4 \cdot 10^{-3}$ & $\geq  3.6 \cdot 10^4$ \\
H$\gamma$ ($c1$) & $2.3 \cdot10^{-3}$ & $3.4 \cdot 10^4$ \\
H$\beta$ ($c2,c1$) & 2.4 - $4.8 \cdot10^{-3}$ & 3.6 - 7.2 $\cdot10^4$  \\
Fe K$\alpha$ narrow ($b$)&  $6.0 \cdot 10^{-2}$ & 8.9 $\cdot10^5$ \\ 
Dust sublimation & $6.0 \cdot 10^{-2}$ & 8.9 $\cdot10^5$ \\
\end{tabular}
\caption{Radial location $r$ (in units of parsec) of the main absorbing/emitting components. References: $a$ \cite{2011MNRAS.414.1965L}; $c1$ \cite{2017ApJ...840...97F}; $c2$ \cite{2000ApJ...542..161P}; $d$ \cite{2019ApJ...879..102P}; $e$ Paper I; $b$ \cite{reeves26}}
\label{tab_radii}
\end{table}

\section{Conclusion \label{sec:Conclusion}}
We applied the time-evolving ionisation code TEPID to a joint XMM-Newton+NuSTAR observation of NGC 4051, with a total observing time of 160 ksec. The observation was divided in 37 time-resolved bin, each $few$ kilosecond long and with  $\approx 2-5\cdot 10^4$ counts in the 0.5-10 kev band. Through the variation of the absorption features in the different spectra, we constrained the temporal evolution of the three Warm Absorbers, whose ionisation follows the intrinsic variation of the ionising continuum. Since the variability timescale depends on the gas density, $t_{eq} \propto 1/ n_e$, we have been able to self-consistently derive $n_e$, and, thus, the radial location $r$ for one WAs and to get upper/lower limits for the other two. The outflows are located at characteristic distances:
\begin{itemize}
\item The Low-Ionisation Phase (LIP) has $log(n_e/cm^{-3}) < 5.5, log(U_0)=-0.03$ and it is likely co-spatial with the outer cold and dusty "torus". The lower limit on $r>2.5 \cdot 10^{-2} pc$ is close to the dust sublimation radius and the FWHM-derived rotational radius of the narrow component of the Fe K$\alpha$ line, which are both at 0.06 $pc$. 
\item The High-Ionisation Phase (HIP) has fully constrained density and location: $log(n_e/cm^{-3}) = 7.0, r=8.2 \cdot 10^{-4} pc$. Its radius lye inside the BLR, which is likely stratified in ionisation and extends between $r \approx 2 \cdot 10^{-4} pc$ (for the He II $\lambda$4686 line) and $r \approx 5 \cdot 10^{-3} pc$ (for the H$\beta$). Similarly, the variability timescale of the hard X-ray reflection component translates into a distance between 0.8 and  $ 1.7 \cdot 10^{-3} pc$.
\item The highest ionisation and velocity component, the HVIP, has $v_{out}=5800$ km s$^{-1}$ (compared with $\approx 500$  km s$^{-1}$ for LIP and HIP), high ionisation and density, $log(U_0)=2.37, log(n_e/cm^{-3}) > 8.7$, and is at $r \leq 4.0 \cdot 10^{-5} pc$ ($r \leq 600 r_G$). The rotationally-derived radius of the broad component of the Fe K$\alpha$ is remarkably similar, $r= 2.3 - 5.0 \cdot 10^{-5} pc$.
\end{itemize}

Basic kinematic considerations rules out a scenario in which the WAs are due to single outflowing shells. Instead, the above findings suggest that LIP, HIP and HVIP are co-located with the bound, virialised structures corotating with the accretion disc and may represent a kind of hotter and steady outflowing phase. In this sense, it is interesting that the O VII He-like triplet resolved in the soft X-ray by \cite{2019ApJ...879..102P} and co-located with the BLR has consistent $r$ with the HIP and higher $n_e$. All these winds have line of sight-projected velocities smaller than the escape velocity at their location: unless their real velocities are much higher (implying vertical, conical-shaped flows) they are bound to fall back on the accretion disc.
Finally, all the three components have very low mass and energy outflow rates and so are likely inefficient in mechanically transferring the accretion-liberated energy at galactic scales, even though they may still heat and transfer metal to the ISM. 

These results demonstrate the power of time-resolved spectroscopy, coupled with time-evolving ionisation modelling. High energy resolution spectrometers with good effective area, such as XRISM/Resolve and the upcoming NewAthena/XIFU and HUBS, finally allow to systematically apply this technique to a huge number of variable and sufficiently bright AGNs, as well as compact sources (see e.g. \citealp{2025Natur.641.1132X,2025A&A...699A.228M,2025arXiv251007615K,reeves26,2026arXiv260416148M}).

\begin{acknowledgements}
AL, FN, LP acknowledge financial support from EU HORIZON-2020 grant “AHEAD2020” (Agreement No. 871158) and from ASI (Italian Space Agency) through the Contract No. 2019-27-HH.0 on Athena. AL, FN acknowledge financial support from PRIN-MUR-2022 grant “Advanced X-ray modeling of black hole winds” (DRAGON; No. PRIN 2022K9N5B4). YK acknowledges support from DGAPAPAPIIT grant IN102023. RS and FN acknowledge financial support from INAF-PRIN grant “A Systematic Study of the largest reservoir of baryons and metals in the Universe: the circumgalactic medium of galaxies” (No. 1.05.01.85.10). RS acknowledges funding from the CAS-ANID grant No. CAS220016.
\end{acknowledgements}

\bibliography{bibliography.bib}
\bibliographystyle{aasjournal}

\appendix
\section{Fit of the time-averaged spectra}
\label{appendix_timeavg}
We fit the time-averaged EPIC-pn, Nustar FPMA and FPMB spectra for Obs1 and Obs2 with the same model as in Eq. \ref{eq:model}, i.e. a continuum made of a black body, a powerlaw and a reflection component, absorbed by  LIP, HIP, HVIP. The only difference is that here the TEPID table models only have the two temporal windows for Obs1 and Obs2, instead of the 37 time-resolved bins. Soft X-ray lines are included as well. Table \ref{tab_timeavg} reports the best-fit values.
We adopt the same strategy as for the time-resolved dataset: the black body temperature is linked between Obs1 and Obs2 and $log(N_H/cm^{-2}), v_{out}$ and the line velocity broadening for the LIP, HIP, HVIP are fixed to the RGS value from Paper I. For simplicity, we do not report such fixed values in the Table. As expected, all the best-fit values are consistent with those from the time-resolved fit but have larger uncertainties. Particularly, $n_e$ for the HIP is not fully constrained and only a lower limit is determined.

\begin{figure*}
\centering
\includegraphics[width=2.\columnwidth]{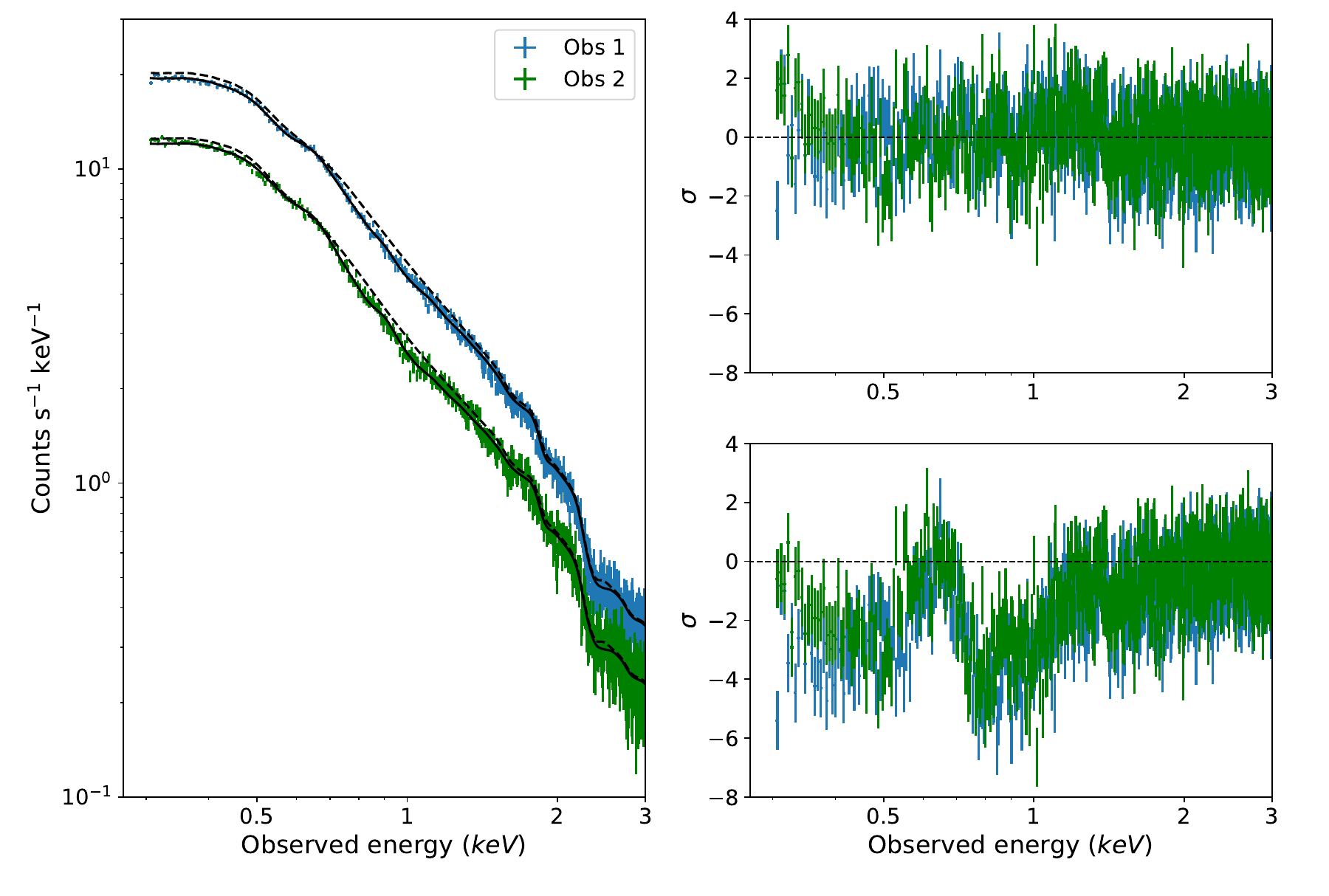}
\caption{Time-averaged spectra for Obs1 and Obs2 (blue and green points, respectively). Left panel shows the data together with the best-fit model (black solid lines). As a reference, we show the models but removing the LIP, HIP, HVIP components as black dashed lines. Right panels show the residuals (in units of $\sigma$) for the complete model (top) and for the model without LIP, HIP, HVIP (bottom). In all the plots the energy axis is restricted to the 0.3-3 keV band, where most of the absorption features are detected.}
\label{fig_timeavg}
\end{figure*}

Fig. \ref{fig_timeavg}, left, shows the EPIC-pn spectra with the best-fit models (black solid line). To visually show their impact, we remove LIP, HIP and HVIP and plot the corresponding models with black dashed lines. Right panels show the residuals, again with and without including the absorbers (top and bottom panels, respectively).

\begin{table}[]
\centering
\begin{tabular}{l|c }
Parameter & Value \\
\hline
\hline
\textbf{norm} (fpma) & $1.14 \pm 0.01$($^l$) \\
\textbf{norm} (fpmb) & $1.14 \pm 0.01$($^l$) \\
\textbf{LIP} \\ 
$\log(U)$ & $-0.3^{+0.1}_{-0.2}$ \\
$\log(n/cm^{-3})$ & $<5.6$ \\
\textbf{HIP} \\ 
$\log(U)$ & $1.64^{+0.03}_{-0.04}$ \\
$\log(n/cm^{-3})$ & $> 5.6$ \\
\textbf{HVIP} \\
$\log(U)$ & $>2.48$ \\
$\log(n/cm^{-3})$ & $>8.7$ \\
\textbf{bb} \\
kT (eV) & $98.9 \pm 0.5$ \\
norm $^a$ (Obs1) & $1.26 \pm 0.03$\\
norm $^a$ (Obs2) & $1.02 \pm 0.02$\\
\textbf{powerlaw} \\
$\Gamma$ (Obs1) & $2.01 \pm 0.01$ \\
$\Gamma$ (Obs2) & $1.87 \pm 0.01$ \\
norm $^{b}$ (Obs1) & $6.10^{+0.07}_{-0.02}$ \\
norm $^{b}$  (Obs2) & $3.41^{+0.02}_{-0.01}$ \\
\textbf{xillver} \\ 
$\Gamma$ (obs1) & $2.00^{+0.03}_{-0.05}$ \\
$\Gamma$ (obs2) & $1.93^{+0.05}_{-0.06}$\\
$A_{fe}$ & $0.73\pm0.05$ \\
norm (obs1) $^c$ & $1.25 \pm 0.08$ \\
norm (obs2) $^c$ & $0.77\pm0.06$ \\
\hline
\hline
$\chi^2/$ deg. of freedom \\
EPIC-pn & 2120.4/2732 (=0.78) \\
FPMA & 445.0/481 (=0.92) \\
FPMB & 379.7/465 (=0.81) \\
total (obs1+obs2) & 2945.1/3710 (=0.79) \\
\end{tabular}
\caption{Best-fit values of the time-averaged EPIC-pn and NuSTAR fit of Obs1 and Obs2. Values are always linked between the three instruments within a given observation, except for the normalisation constants. Fit statistics (at the bottom) are given separately for each dataset and for the total. All values are linked between Obs1 and Obs2 except when stated. Normalisations are in units of: $^a$, $10^{-4} L_{39}/D_{10}$, where the ratio is between the luminosity in units of $10^{39} erg/s$ and the distance in units of 10 kpc; $^b$, $10^{-3}$ ph/keV/cm$^2$/s; $^c$, $10^{-4}$.}
\label{tab_timeavg}
\end{table}

\end{document}